\documentclass[sigconf]{acmart}

\usepackage{bm}
\usepackage{algorithm}
\usepackage{algorithmic}
\usepackage{bbm}
\newcommand{\figscale}{0.127}
\newcommand{\figscalea}{0.15}

\AtBeginDocument{%
  \providecommand\BibTeX{{%
    \normalfont B\kern-0.5em{\scshape i\kern-0.25em b}\kern-0.8em\TeX}}}

\setcopyright{acmlicensed}
\copyrightyear{2020}
\acmYear{2020}
\acmDOI{10.1145/3394486.3403116}

\acmConference[KDD '20]{Proceedings of the 26th ACM SIGKDD Conference on Knowledge Discovery and Data Mining}{August 23--27, 2020}{Virtual Event, CA, USA}
\acmBooktitle{Proceedings of the 26th ACM SIGKDD Conference on Knowledge Discovery and Data Mining (KDD '20), August 23--27, 2020, Virtual Event, CA, USA}
\acmPrice{15.00}
\acmISBN{978-1-4503-7998-4/20/08}

\begin{document}
\fancyhead{}

\title{Estimating Properties of Social Networks via Random Walk considering Private Nodes}

\author{Kazuki Nakajima}
\affiliation{\institution{Tokyo Institute of Technology}}
\email{nakajima.k.an@m.titech.ac.jp}

\author{Kazuyuki Shudo}
\affiliation{\institution{Tokyo Institute of Technology}}
\email{shudo@is.titech.ac.jp}


\begin{abstract}
Accurately analyzing graph properties of social networks is a challenging task because of access limitations to the graph data. 
To address this challenge, several algorithms to obtain unbiased estimates of properties from few samples via a random walk have been studied. 
However, existing algorithms do not consider {\it private nodes} who hide their neighbors in real social networks, leading to some practical problems. 
Here we design random walk-based algorithms to accurately estimate properties without any problems caused by private nodes.
First, we design a random walk-based sampling algorithm that comprises the neighbor selection to obtain samples having the Markov property and the calculation of weights for each sample to correct the sampling bias. 
Further, for two graph property estimators, we propose the weighting methods to reduce not only the sampling bias but also estimation errors due to private nodes.
The proposed algorithms improve the estimation accuracy of the existing algorithms by up to 92.6\% on real-world datasets.
\end{abstract}

\ccsdesc[500]{General and reference~Estimation}
\ccsdesc[500]{Mathematics of computing~Graph algorithms}

\keywords{Social networks; Estimation; Sampling; Random walk; Private nodes}


\maketitle

\section{Introduction}
Online social networks (OSNs) have been primarily studied to understand the nature of global social structures, such as human connections and behaviors \cite{ahn, gjoka_practical, gjoka_walking, kwak, mislove, wilson}. 
A widely used approach to understand the structure is to analyze various graph properties, such as the number of nodes (size), the average degree, and the degree distribution. However, an accurate analysis of social graphs is a challenging task because of access limitations to the graph data. 
Although some studies used complete graph data \cite{ahn, backstrom, myers}, all the data is typically not available to researchers. 
In practical scenarios, we sample a part of graph data through the application programming interfaces (APIs) to estimate properties \cite{gjoka_practical, gjoka_walking}. 
We can obtain unbiased estimates if users are randomly sampled; however, it is frequently difficult to randomly assign user IDs to the APIs \cite{dasgupta, katzir_nodecc}.

Several algorithms \cite{chen, dasgupta, katzir_nodecc, katzir_node, nakajima, wang} based on the {\it re-weighted random walk} scheme \cite{gjoka_practical, gjoka_walking} have been studied to address this challenge. Many OSNs have provided APIs to return the neighbor data of a queried user. By repeatedly utilizing this function for a randomly selected neighbor, we obtain samples with the Markov property via a random walk on a social network. 
Then, we obtain unbiased estimates of properties by re-weighting each sample to correct the sampling bias derived from the Markov chain analysis. 
Previous studies have focused on accurate estimates with a small number of queries because {\it APIs are typically rate-limited} \cite{dasgupta, katzir_nodecc}.

However, previous studies have ignored {\it private users} who do not provide their neighbor data even if queried. 
Although previous studies assume a network comprising only {\it public users} who publish their neighbor data, a certain percentage of private users exist in real social networks. 
For example, private users were about 27\% on Facebook \cite{catanese} and 34\% on Pokec \cite{takac}, which is an OSN in Slovakia. 

Private users cause some problems for existing algorithms. 
For example, how do we deal with private users in a random walk sampling? 
If private users are sampled, we need to handle an exception wherein the neighbor data cannot be obtained by approaches such as returning to the public user sampled previously. 
However, if such an exceptional process is performed, the Markov property of a sample sequence cannot be guaranteed. 
This can prevent us from obtaining unbiased estimates of properties. 
There is another serious problem. 
A fast solution of problems in sampling is to not include private users to the sample sequence; however, by not sampling them, the sampling bias and weighting for each sample are different from the assumptions of existing algorithms. 
Thus, the existing algorithms cause estimation errors due to private users. 

We aim to design re-weighted random walk algorithms considering private users to accurately estimate properties without any problems caused by private users. To the best of our knowledge, this study is the first to focus on the effect of private users on random walk-based estimators. First, we make three assumptions and define two access models ({\it the ideal model} and {\it the hidden privacy model}), based on previous studies and our observations of real social networks (Section \ref{preliminaries}). 
Then, we design sampling and estimation algorithms based on these assumptions and access models.

This study has two main contributions. The first contribution is to enable us to {\it successfully perform re-weighted random walk-based algorithms in real social networks including private users} (Section \ref{sampling}). We discuss the neighbor selection in a random walk by considering private users and derive the sampling bias of each user. Then, for each access model, we describe the calculation of the weights to correct the sampling bias; particularly for the hidden privacy model, we propose a method to approximate weights with much fewer queries than exact calculation.

The second contribution is to enable us to {\it accurately estimate the size and average degree of a whole social graph including private nodes} (Section \ref{estimation}). Existing algorithms \cite{dasgupta, katzir_nodecc, katzir_node} cause errors due to private users because the conventional weighting aims only to correct the sampling bias. We propose weighting methods to reduce both sampling bias and errors due to private users. Moreover, we theoretically explain that estimates obtained using the proposed weighting converge to values with smaller expected errors than the previous weighting. 

Finally, we validate the theoretical claims and effectiveness of the proposed algorithms on extensive experiments using real social network datasets (Section \ref{experiments}). It is important to evaluate the performance when the assumptions are not satisfied because the proposed algorithms are designed on some simplified assumptions. We empirically show that the proposed algorithms perform well on realistic datasets, such as a 92.6\% improvement in the estimation accuracy on the Pokec dataset including real private users.

\section{Preliminaries}\label{preliminaries}
\subsection{Definitions and Notations}
We represent a social network as a connected and undirected graph, $G = (V, E)$, where $V = \{v_1, ..., v_n\}$ denotes the set of nodes (users), and $E$ denotes the set of edges (friendship). Let $d_i$ denote the degree (number of neighbors) of $v_i$ and $D = \sum_{v_i \in V} d_i$ denote the sum of degrees. We define the average degree of $G$ as $d_{avg} = \frac{D}{n}$. Each node, $v_i$, has a privacy label, $l_{pri}(i) \in \{public, private\}$. The set of privacy labels is denoted by $\mathcal{L}_{pri} = \{l_{pri}(i)\}_{v_i \in V}$. 

We refer to connected subgraphs that consist of public nodes on $G$ as {\it public-clusters}. Let $\{C_i\}$ denote a set of public-clusters and $C^* = (V^*, E^*)$ denote the largest public-cluster. Let $n^*$ denote the number of nodes in $C^*$, and  $d_i^* = |\{v_j \in V^* | (v_i,v_j) \in E^* \}|$ denote the {\it public-degree} (the number of neighbors that publish their own neighbors) of a node $v_i \in V^*$. Let $D^* = \sum_{v_i \in V^*} d_i^*$ denote the sum of public-degrees. We define the average degree of $C^*$ as $d_{avg}^* = \frac{D^*}{n^*}$. Let $\bm{1}_{V^*}(v_i)$ denote a function that returns 1 if $v_i \in V^*$, otherwise 0.

\begin{figure}[t]
        \begin{center}
          \includegraphics[scale=0.135]{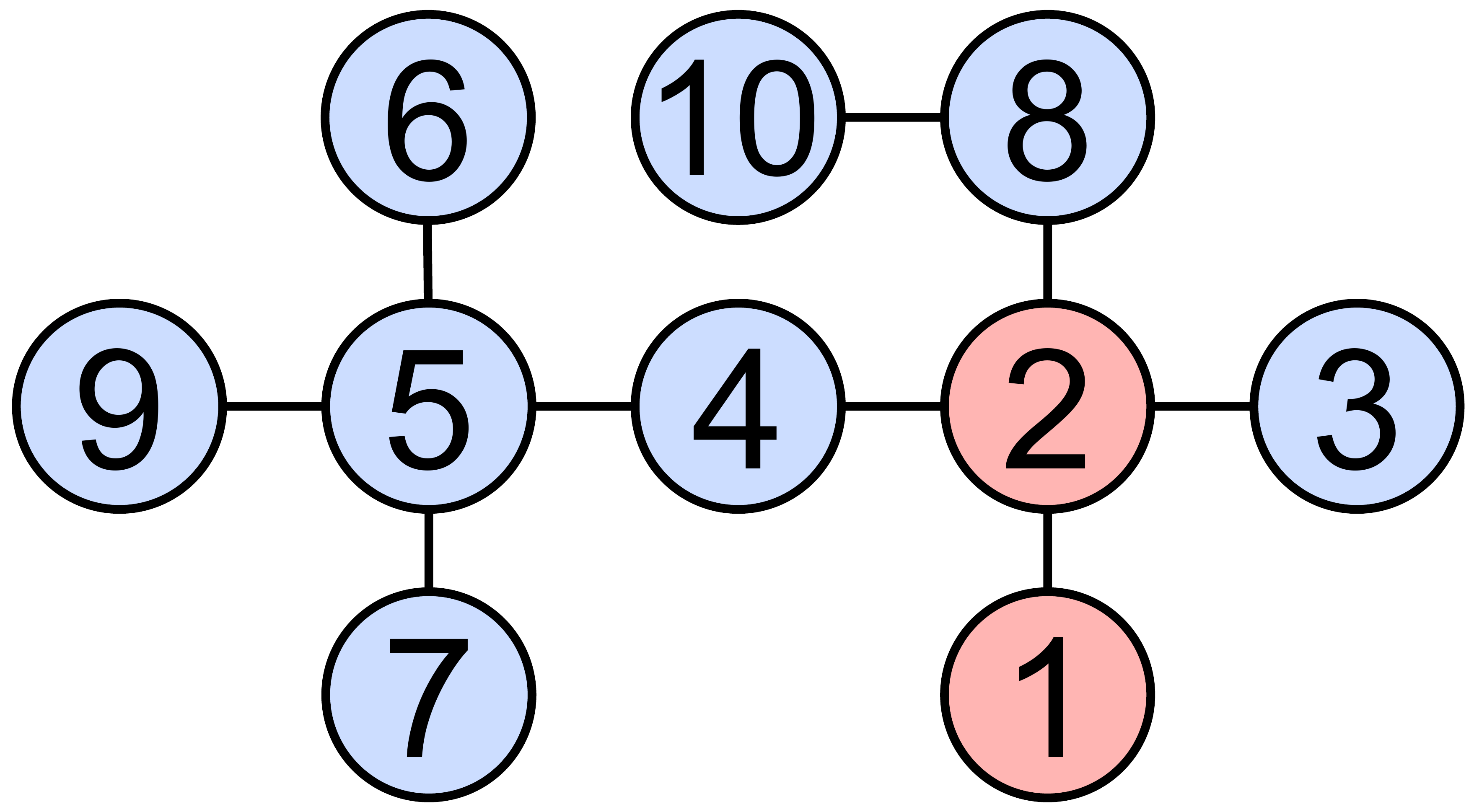}
        \end{center}
      \caption{An example of a graph with privacy labels. Here, nodes 1 and 2 are private and others are public.}
      \label{graph}
\end{figure}

{\bf Example:} Let $v_i = i\ (1 \leq i \leq 10)$ in Figure \ref{graph}. There are three public-clusters, $C_1$, $C_2$ and $C_3$, and $C_1$ is the largest public-cluster:
\begin{itemize}
  \item $C_1 = (\{4, 5, 6, 7, 9\},\{(4, 5), (5, 6), (5, 7), (5, 9)\})$
  \item $C_2 = (\{8, 10\}, \{(8, 10)\})$
  \item $C_3 = (\{3\}, \{\})$.
\end{itemize}
Also, it holds $n^* = 5, d_4^* = 1, d_5^* = 4, d_6^*=1, d_7^*=1, d_9^*=1$.

\subsection{Assumptions}
We design algorithms based on the following three assumptions.

\begin{enumerate}
\item {\it The indices of all neighbors, including private neighbors, of a queried public node are obtained.} Facebook as of the previous study \cite{gjoka_practical} satisfied this assumption. We have confirmed that Twitter as of June 2020 satisfies this assumption \cite{twitter_api_followers, twitter_api_friends}. \label{assumption_public}

\item {\it Each node independently becomes private with probability $p$, otherwise public, where $0 \leq p \leq 1$.} 
We assume that each node independently hides neighbors with a certain probability\footnote{Intuitively, private nodes tend to have low degrees under Assumption 2, because the degree distribution of social networks is typically biased to low degrees \cite{ahn,gjoka_practical, gjoka_walking,kwak,mislove}.}. \label{assumption_private}

\item {\it The seed of a random walk is on the largest public-cluster.} We do not consider the number of queries performed to select the seed from nodes on the largest public-cluster, because we consider that this number is sufficiently small.\label{assumption_seed}
\end{enumerate}

\subsection{Access Models}
We do not assume the acquisition of all the graph data and their random access. We assume that a graph $G$ is static. 

If we query a public node, the neighbor data are provided. If a private node is queried, the neighbor data cannot be obtained\footnote{The output may be an error message or an empty set.}. We consider two models for the neighbor data that are provided when a public node, $v_i$, is queried: ideal model and hidden privacy model.

{\bf The ideal model: }{\it Indices (IDs) and privacy labels of all neighbors of $v_i$ are obtained.} For example, when a public node 4 is queried in Figure \ref{graph}, we obtain the set $\{(2, private), (5, public)\}$. Facebook as of the previous study \cite{gjoka_practical} corresponds to this access model. 

{\bf The hidden privacy model: }{\it Even though ids of all neighbors of $v_i$ are obtained, the privacy labels are not.} For example, when a public node 4 is queried in Figure \ref{graph}, we obtain the set $\{2, 5\}$. 
Twitter API as of June 2020 corresponds to this model \cite{twitter_api_followers, twitter_api_friends}.

\subsection{Markov Chain Basics}
We introduce the basics of a Markov chain for theoretical analysis of random walk-based estimators. 
First, we describe the stationary distribution of a Markov chain, which is needed to derive the sampling bias. Let $\bm{P} = \{P_{i,j}\}_{i,j \in S}$ denote the transition probability matrix of a Markov chain on the state space, $S$. If $\pi_j  = \sum_{i \in S} {\pi_i P_{i,j}}$ holds for all $j \in S$, $\bm{\pi} = \{\pi_i\}_{i \in S}$ is the stationary distribution of $\bm{P}$. The following theorem holds for the stationary distribution $\bm{\pi}$ of $\bm{P}$:
\begin{theorem}\cite{levin}\label{ergodic}
If $\bm{P}$ is ergodic (i.e., irreducible and aperiodic), the stationary distribution, $\bm{\pi}$, uniquely exists.
\end{theorem}

Then, we review the strong law of large numbers for a Markov chain, which is needed to ensure that the sample average converges almost surely to its expected value:
\begin{theorem}\cite{chen, lee_nonback}\label{SLLN}
Let $\{X_k\}$ be an ergodic Markov chain with the stationary distribution $\bm{\pi}$ on the finite state space $S$. For any function $f: S \to \mathbb{R}$, $\frac{1}{r} \sum_{k=1}^r f(X_k)$ converges to $\sum_{i \in S} \pi_i f(i)$ almost surely as $r \to \infty$ regardless of the initial distribution of the chain.
\end{theorem}

\section{Random Walk Sampling}\label{sampling}
We design a random walk-based sampling algorithm that consists of the neighbor selection to obtain samples having the Markov property and the calculation of weights to correct the sampling bias.
The proofs in this section are shown in the supplement.

\subsection{Transition Neighbor Selection}\label{neighbor}
In a simple random walk, we move to a randomly selected neighbor. 
If private nodes are sampled, practical problems on estimating properties occur, because their neighbor data are unclear. 
For example, we cannot simply continue the algorithm, and it is difficult to correct the sampling bias of sampled private nodes. 

We believe that it is appropriate to move to a public neighbor randomly selected as Gjoka et al. \cite{gjoka_practical} performed on Facebook\footnote{There is no detailed discussion of the effect of private nodes on re-weighted random walk algorithms \cite{gjoka_practical}. They focused on only the part comprising public nodes.}. 
We design the algorithm to obtain an index sequence of $r$ sampled nodes, denoted by $\{x_k\}_{k=1}^r$, as follows. 
First, we select a seed, $v_{x_1} \in C^*$, as the first sample. 
Then, for $1 \leq k \leq r-1$, we randomly select a node, $u$, from neighbors of the $k$-th sample, $v_{x_k}$. 
If $u$ is public, we move to $u$ as next sample, $v_{x_{k+1}}$; otherwise, we randomly reselect a neighbor of $v_{x_k}$.  
In the ideal model, we check if a selected neighbor $u$ is public without an additional query. 
In the hidden privacy model, we judge the privacy label of $u$ by additionally querying $u$.

\subsection{Sampling Bias}
We derive the sampling bias caused by the designed random walk. 
Let the probability that an event $A$ will occur be denoted by $Pr[A]$. We define the distribution induced by the sequence of sampled indices as $\bm{\pi}_{r} = (Pr[x_r = i])_{v_i \in V}$. 

We show that each node on the largest public-cluster is sampled in proportion to the public-degree via the designed random walk.
\begin{lemma}\label{distribution}
$\bm{\pi}_{r}$ converges to the vector $\bm{\pi} = (p_i \bm{1}_{V^*}(v_i))_{v_i \in V}$ after many steps of the designed random walk, where $p_i = \frac{d_i^*}{D^*}$.
\end{lemma}

\subsection{Public-Degree Calculation}\label{calc_pubd}
We need to calculate the public-degree of each sample to correct the sampling bias due to the public-degree.

In the ideal model, we can exactly calculate each public-degree without additional query because the privacy labels of all the neighbors are obtained. Conversely, in the hidden privacy model, the exact calculation needs a considerable number of additional queries to obtain privacy labels of all neighbors for each sample.

For the hidden privacy model, we propose a method to approximate each public-degree without additional queries by utilizing the history of neighbor selections of the designed random walk.
We record two values for each sample $v_{x_k}$: the total number of successful public neighbor selections, $a_{x_k}$, and that of neighbor selections, $b_{x_k}$.
After sampling, each approximate value, $\hat{d}_{x_k}^*$, is calculated as
\begin{align*}
  \hat{d}_{x_k}^* \triangleq d_{x_k} \frac{a_{x_k}}{b_{x_k}}.
\end{align*}
The pseudo-code of the designed random walk with the proposed approximation of public-degrees is shown in the supplement.

We ensure the convergence of the approximated public-degree after many steps of the designed random walk. 
\begin{lemma}\label{the_pubd_exp}
$\hat{d}_{x_k}^*$ converges to $d_{x_k}^*$ for each sample, $v_{x_k}$, after many steps of the designed random walk.
\end{lemma}

Then, we explain that the proposed method performs much fewer queries than the exact method that exactly calculates the public-degree of each sample. 
Here, for simplicity, we do not consider the save of the indices of nodes queried once. $Q(k)$ and $\hat{Q}(k)$ denote the number of queries that are performed at the $k$-th sample by the exact and proposed methods, respectively. 
Let $Q = \frac{1}{r} \sum_{k=1}^r Q(k)$ and $\hat{Q} = \frac{1}{r} \sum_{k=1}^r \hat{Q}(k)$ denote the ratio of the number of queries to the sample size of the exact and proposed methods, respectively.

\begin{lemma}\label{query_efficiency}
The expectations of $Q$ and $\hat{Q}$ are $\frac{\sum_{v_i \in V^*} d_i^* d_i}{D^*}$ and $\frac{\sum_{v_i \in V^*} d_i}{D^*}$, respectively. 
\end{lemma}

Lemma \ref{query_efficiency} implies that the proposed method performs much fewer queries than the exact method because $\sum_{v_i \in V^*} d_i$ is intuitively much smaller than $\sum_{v_i \in V^*} d_i^* d_i$ in real social networks.

\section{Properties Estimation}\label{estimation}
We design estimators for the size and the average degree via a random walk designed in Section \ref{sampling}. 
We first introduce the existing estimators and then propose estimators with improved weighting.
The proofs in this section are shown in the supplement.

\subsection{Overview}
Here we address a novel problem that the convergence values of estimators have errors caused by private nodes. 
The conventional estimators perform the weighting using only the public-degree to correct the sampling bias; consequently, the estimates converge to the properties of the largest public-cluster. 
When a graph comprises only public nodes as have been assumed in previous studies, the convergence values are equal to the properties of the original graph. 
However, when there are private nodes on a graph, the convergence values typically have errors due to private nodes, or more specifically, the value of $p$. 

It is unclear how to reduce the errors due to the value of $p$. If we know the exact value of $p$, we can easily correct the errors; however, this is usually unknown. Moreover, we cannot apply existing methods to estimate the value of $p$ based on privacy labels of samples (Section 3.C.3 in \cite{gjoka_practical}, Section 3.2 in \cite{katzir_node} and Section 4.2.3 in \cite{ribeiro}), because private nodes are not included in samples.

We propose the weighting methods combining the public-degree and degree to reduce the errors of convergence values due to the value of $p$. The proposed methods are based on the fact that the public-degree follows the binomial distribution with parameters of the degree and $1-p$ regarding the set of privacy labels under Assumption \ref{assumption_private}. Proposed estimators provide the same convergence values when $p = 0$ and convergence values with smaller expected errors when $p > 0$ as compared with the existing estimators.

\subsection{Size Estimation}\label{size_estimation}
{\bf Existing estimator: }The node collision (NC) algorithm is an effective estimator for the size \cite{katzir_nodecc,katzir_node}. 
This algorithm examines sample pairs whose ordinal numbers in the sample sequence are more than a threshold, $m$, apart. 
Sample pairs whose ordinal numbers in the sample sequence are sufficiently apart can be regarded as being independently sampled \cite{katzir_nodecc}.

Let $I = \{(k,l)\ |\ m \leq |k-l| \land 1 \leq k,l \leq r\}$ denote the set of integer pairs that are between $1$ and $r$ and at least $m$ away. Let $\phi_{k,l}$ denote a variable which returns 1 if $x_k = x_l$ (this is called a collision), and otherwise, 0. The average of the number of collisions $\Phi^{NC}_{n}$, the average of the weights to correct the sampling bias $\Psi^{NC}_{n}$, and a size estimate $n^{NC}$ obtained by NC algorithm are as follows:
\begin{align*}
\Phi^{NC}_{n} = \frac{1}{|I|}\sum_{(k,l) \in I} \phi_{k,l},\ \ \ 
\Psi^{NC}_{n} = \frac{1}{|I|}\sum_{(k,l) \in I} \frac{d_{x_k}^*}{d_{x_l}^*},\ \ \ 
n^{NC} \triangleq \frac{\Psi^{NC}_{n}}{\Phi^{NC}_{n}}
\end{align*}

It holds the following lemma expanded from the previous studies \cite{katzir_nodecc,katzir_node} which assume graph with no private nodes. 
\begin{lemma}\label{the_size_exp_naive}
$n^{NC}$ converges to size of the largest public-cluster, $n^*$.
\end{lemma}

We focus on the expectation of $n^*$ regarding the set of privacy labels, $\mathcal{L}_{pri}$, to quantify the error between $n^*$ and $n$. 
$E_{pri}[X]$ denotes the expectation of a random variable $X$ regarding $\mathcal{L}_{pri}$. 
We approximate the expectation of $n^*$ regarding $\mathcal{L}_{pri}$ under the condition that all the public nodes belong to the largest public-cluster. 
Under this condition, it holds that $Pr[v_i \in V^*] = Pr [l_{pri}(i) = public] = 1-p$. 

The following lemma holds for the convergence value $n^*$:
\begin{lemma}\label{error_size_naive}
Under the condition that all the public nodes belong to the largest public-cluster, we have
\begin{align*}
  E_{pri}[n^*] = (1-p)n.
\end{align*}
\end{lemma}

Lemma \ref{error_size_naive} implies that an expected relative error of the convergence value in the existing estimator is $1-p$.

{\bf Proposed estimator: }We improve the weighting for each sample pair to reduce the convergence error. We define the average of the proposed weights as $\hat{\Psi}_{n}$, then our estimation value as $\hat{n}$:
\begin{align*}
\hat{\Psi}_{n} = \frac{1}{|I|}\sum_{(k,l) \in I} \frac{d_{x_k}}{d_{x_l}^*},\ \ \ 
\hat{n} \triangleq \frac{\hat{\Psi}_{n}}{\Phi^{NC}_{n}}.
\end{align*}

It holds the following lemma regarding the convergence values of the proposed estimator. 
\begin{lemma}\label{the_size_exp_ours}
$\hat{n}$ converges to $\tilde{n} = n^* \frac{\sum_{v_i \in V^*} d_i^* d_i}{\sum_{v_i \in V^*} (d_i^*)^2}$.
\end{lemma}

When there are no private nodes on $G$, the following proposition holds for each estimate and convergence value.

\begin{proposition}\label{p0_size}
When there are no private nodes on $G$, the two estimation values, $n^{NC}$ and $\hat{n}$, are equal, and the two convergence values, $n^*$ and $\tilde{n}$, are equal to the true value, $n$.
\end{proposition}

We clarify our convergence value, $\tilde{n}$, has the smaller expected error for any values of $p$ than the previous convergence value, $n^*$. 
First, we show the following lemma:
\begin{lemma}\label{pubd_exp}
For any node $v_i \in V^*$, we have
\begin{align*}
  E_{pri}[d_i^*] &= (1 - p)d_i, \\
  E_{pri}\left[(d_i^*)^2 \right] &= (1 - p)d_i[(1-p)d_i+p].
\end{align*}
\end{lemma}

Then, we approximate the expectation of $\tilde{n}$ regarding $\mathcal{L}_{pri}$ by deriving each expectation of the denominator and numerator of $\tilde{n}$.
\begin{theorem}\label{error_size_ours}
Let $\alpha_p = \frac{(1-p) \sum_{v_i \in V} {(d_i)^2}}{\sum_{v_i \in V} d_i [(1-p)d_i + p]}$. Under the condition that all the public nodes belong to the largest public-cluster, we have
\begin{align*}
  E_{pri}[\tilde{n}] \approx \frac{E_{pri}[n^* ] E_{pri}[\sum_{v_i \in V^*} d_i^* d_i]}{E_{pri}[\sum_{v_i \in V^*} (d_i^*)^2]} = \alpha_p n.
\end{align*}
\end{theorem}

Theorem \ref{error_size_ours} implies that our convergence value is almost equal to the entire size because the coefficient $\alpha_p$ is empirically almost equal to 1 for various values of $p$ in real social networks.

The following corollary claims that the proposed estimator improves an expected convergence error for any values of $p$. 

\begin{corollary}\label{relative_size}
Under the condition that all the public nodes belong to the largest public-cluster, we have
\begin{align*}
  \left|n - \frac{E_{pri}[n^*]E_{pri}[\sum_{v_i \in V^*} d_i^* d_i]}{E_{pri}[\sum_{v_i \in V^*} (d_i^*)^2]}\right| \leq \left|n - E_{pri}[n^*]\right|.
\end{align*}
\end{corollary}

\subsection{Average Degree Estimation}
{\bf Existing estimator: }The Smooth algorithm \cite{dasgupta} takes a rough estimate of the average degree, $c$, as an input and returns a more accurate estimate. Herein, we let $c$ to be 0 because Dasgupta et al. concluded that small values, 0 or 1, are desirable \cite{dasgupta}. 

An estimate of the average degree obtained using the Smooth algorithm, $d_{avg}^{Smooth}$, is defined as follows:
\begin{align*}
d_{avg}^{Smooth} \triangleq \frac{r}{\sum_{k=1}^r {\frac{1}{d_{x_k}^*}}}.
\end{align*}

It holds the following lemma derived from the previous study \cite{dasgupta} which assumes a graph with no private nodes.

\begin{lemma}\label{the_aved_exp_naive}
$d_{avg}^{Smooth}$ converges to the average degree of the largest public-cluster, $d_{avg}^*$.
\end{lemma}

Then, we approximate the expectation of $d_{avg}^*$ regarding $\mathcal{L}_{pri}$.
\begin{lemma}\label{error_aved_naive}
Under the condition that all the public nodes belong to the largest public-cluster, we have
\begin{align*}
  E_{pri}[d_{avg}^*] \approx \frac{E_{pri}[D^*]}{E_{pri}[n^*]} = (1-p)d_{avg}.
\end{align*}
\end{lemma}

Lemma \ref{error_aved_naive} implies that an expected relative error of the convergence value in the existing estimator is $1-p$.

{\bf Proposed estimator: }We improve the weighting to reduce a convergence error due to the value of $p$. We define our estimate as
\begin{align*}
\hat{d}_{avg} \triangleq \frac{r}{\sum_{k=1}^r {\frac{1}{d_{x_k}}}}.
\end{align*}

The following lemma regarding our estimate holds:
\begin{lemma}\label{the_aved_exp_ours}
$\hat{d}_{avg}$ converges to $\tilde{d}_{avg} = \frac{D^*}{\sum_{v_i \in V^*} \frac{d_i^*}{d_i}}$.
\end{lemma}

The following proposition holds as well as Proposition \ref{p0_size}.
\begin{proposition}\label{p0_aved}
When there are no private nodes on $G$, two estimates, $d_{avg}^{Smooth}$ and $\hat{d}_{avg}$, are equal, and two convergence values, $d_{avg}^*$ and $\tilde{d}_{avg}$, are equal to the true value, $d_{avg}$.
\end{proposition}

The following theorem and corollary claim that the proposed estimator improves an expected error of the convergence value.

\begin{figure*}[t]
  \begin{minipage}{0.24\hsize}
        \begin{center}
        \includegraphics[scale = \figscalea]{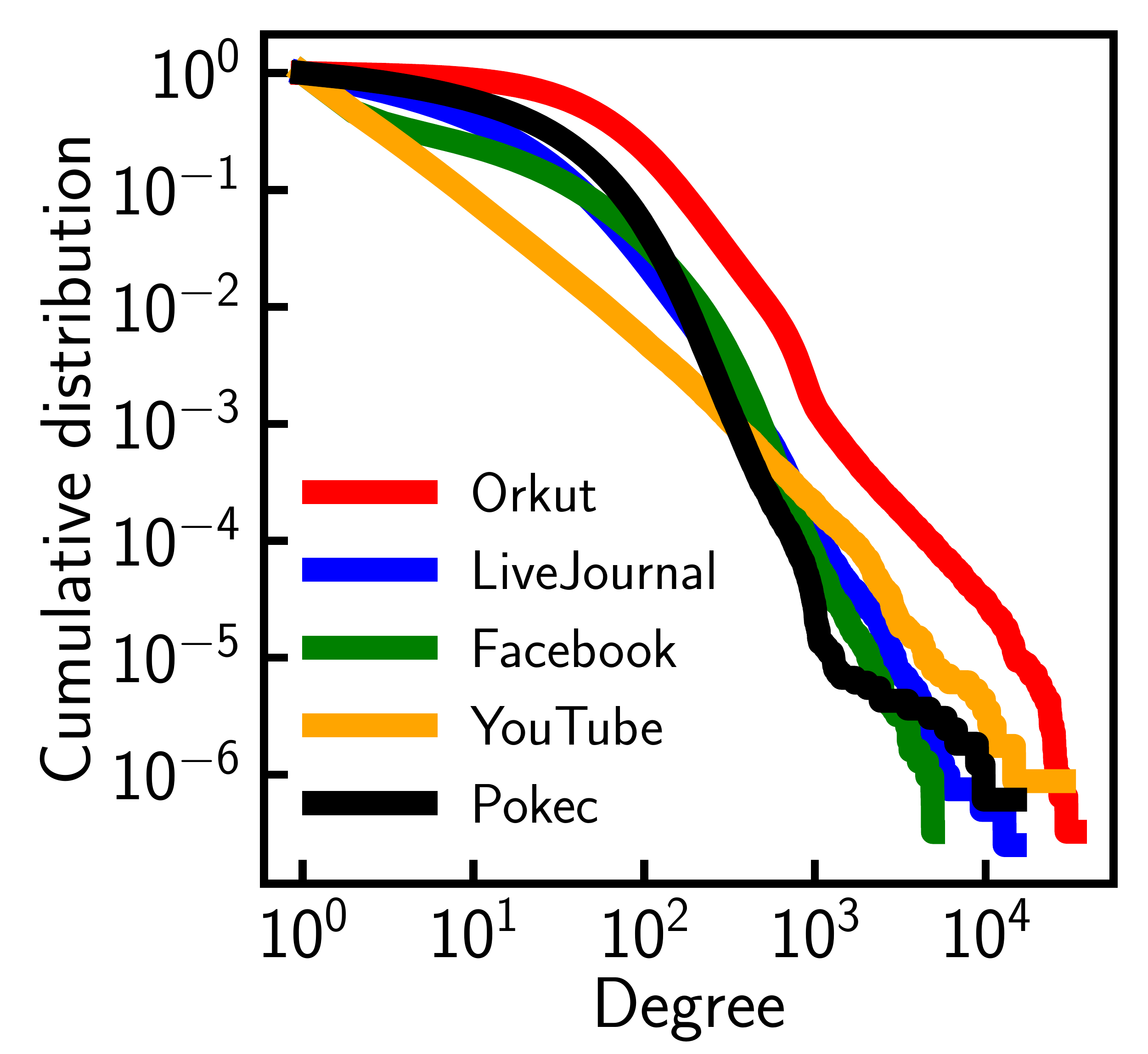}
        \end{center}
        \vspace{-1mm}
        \caption{Cumulative degree distributions.}
        \label{degree_distribution}
        \vspace{-1mm}
  \end{minipage}
  \begin{minipage}{0.24\hsize}
        \begin{center}
          \includegraphics[scale=\figscalea]{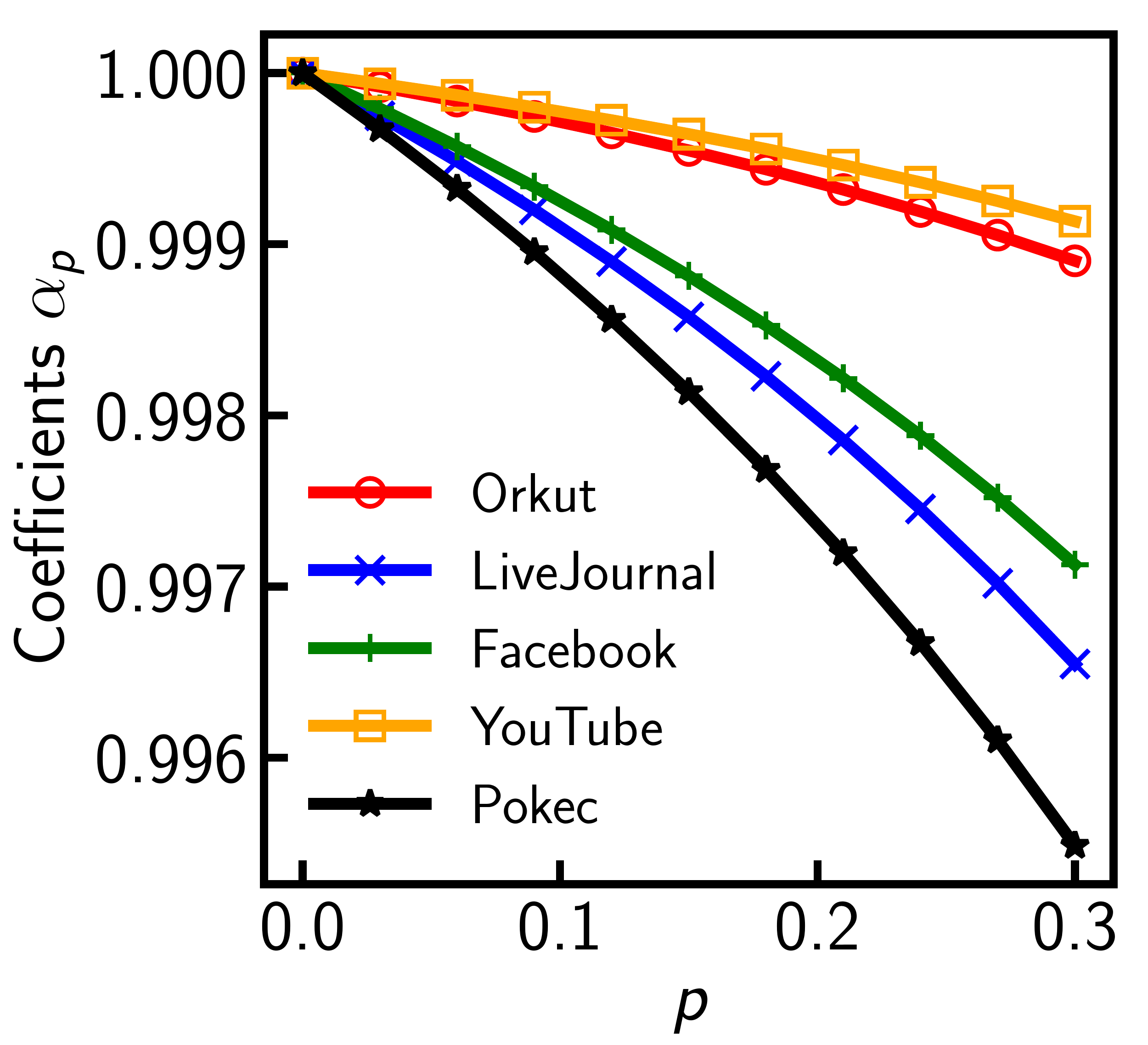}
        \end{center}
        \vspace{-1mm}
        \caption{Coefficients $\alpha_p$.}
        \vspace{3.5mm}
      \label{alpha_p}
      \vspace{-1mm}
  \end{minipage}
  \hspace{2mm}
  \begin{minipage}{0.24\hsize}
        \begin{center}
          \includegraphics[scale=\figscalea]{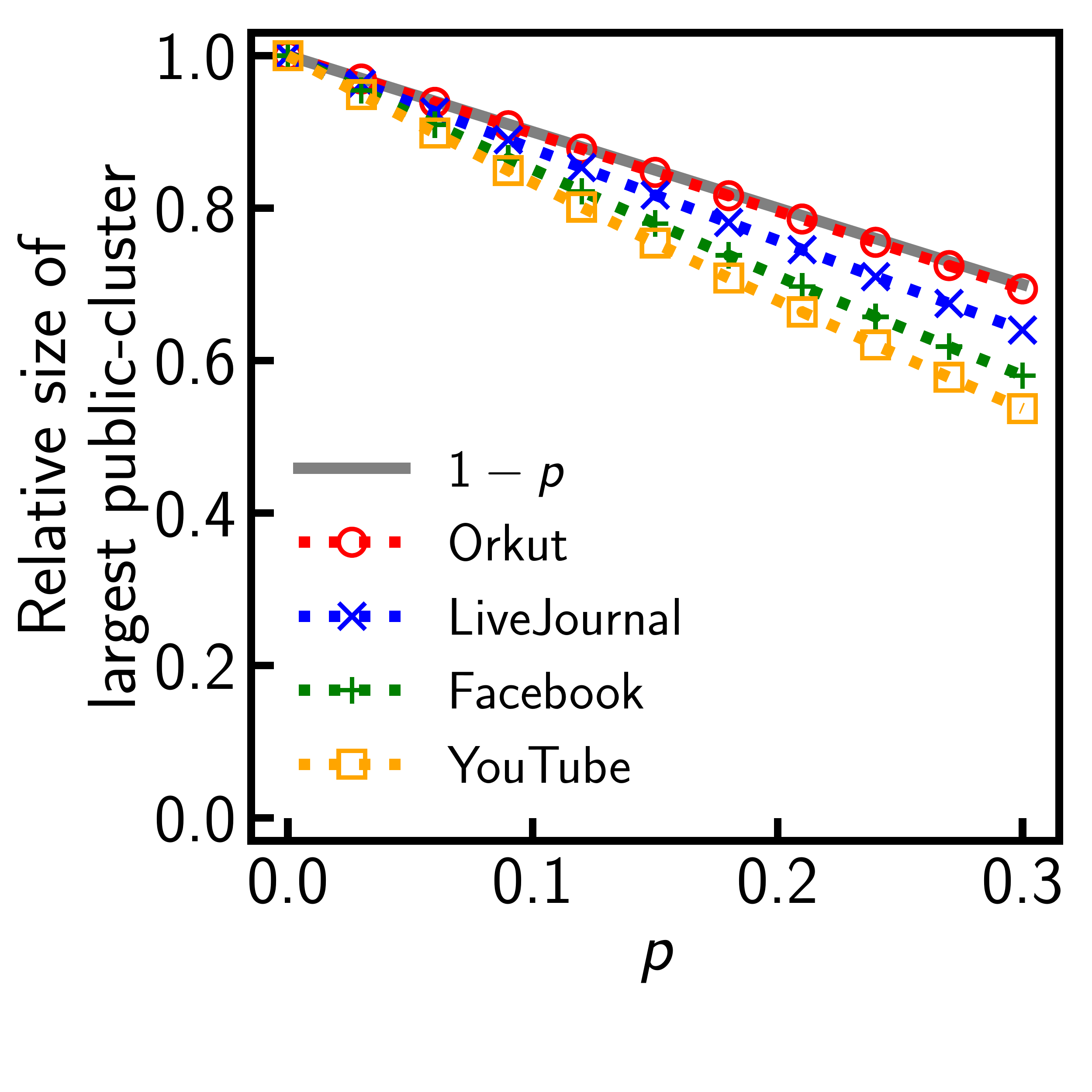}
        \end{center}
      \vspace{-1mm}
      \caption{Relative size of the largest public-cluster.}
      \label{cluster_size}
      \vspace{-1mm}
  \end{minipage}
  \hspace{1mm}
  \begin{minipage}{0.24\hsize}
      \begin{center}
          \includegraphics[scale=\figscalea]{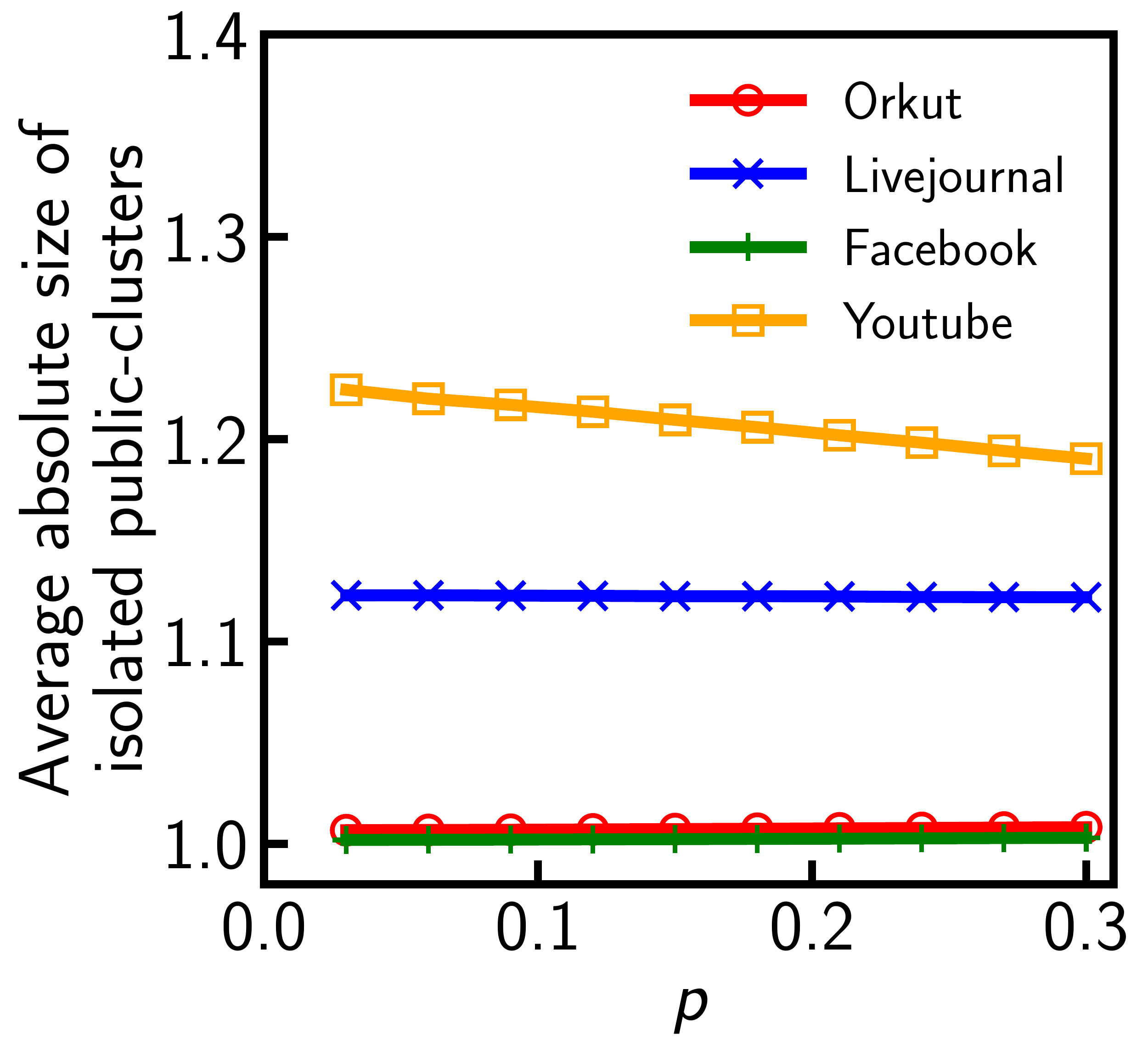}
        \end{center}
        \vspace{-1mm}
        \caption{Average absolute size of isolated public-clusters.}
      \label{isolated_size}
      \vspace{-1mm}
  \end{minipage}
\end{figure*}

\begin{theorem}\label{error_aved_ours}
Under the condition that all the public nodes belong to the largest public-cluster, we have
\begin{align*}
  E_{pri}[\tilde{d}_{avg}] \approx \frac{E_{pri}[D^*]}{E_{pri} \left[\sum_{v_i \in V^*} \frac{d_i^*}{d_i} \right]} = d_{avg}.
\end{align*}
\end{theorem}

\begin{corollary}\label{relative_aved}
Under the condition that all the public nodes belong to the largest public-cluster, we have
\begin{align*}
  \left|d_{avg} - \frac{E_{pri}[D^*]}{E_{pri}\left[\sum_{v_i \in V^*} \frac{d_i^*}{d_i}\right]} \right| \leq \left|d_{avg} - \frac{E_{pri}[D^*]}{E_{pri}[n^*]}\right|.
\end{align*}
\end{corollary}

\subsection{Estimation in the Hidden Privacy Model}
In the hidden privacy model, we calculate each estimate using the approximated public-degree, $\hat{d}_{x_k}^*$. Even in this model, Lemmas \ref{the_size_exp_naive}, \ref{the_size_exp_ours}, \ref {the_aved_exp_naive}, and \ref {the_aved_exp_ours} hold because of Lemma \ref{the_pubd_exp}.

\section{Experiments}\label{experiments}
We evaluate the proposed algorithms using real social network datasets. We aim to answer the following questions:
\begin{enumerate}
  \item Do the proposed estimators improve the estimation accuracy of the size and average degree for various probabilities of $p$? (Section \ref{performance})
  \item Do the proposed estimators perform acceptably in real-world datasets including real private users? (Section \ref{realdataset})
  \item How does the proposed public-degree calculation for the hidden privacy model affect the estimation accuracy and number of queries? (Section \ref{effect_pubd})
  \item Is the number of queries performed in the seed selection due to private nodes is small? (Section \ref{seed})
\end{enumerate}

\subsection{Experimental Setup} \label{setup}
{\bf Datasets:} We use datasets of known size and average degree, YouTube, Pokec, Orkut, Facebook, and LiveJournal. 
For these five datasets, we focus on undirected and connected graphs by the pre-processing: (1) removing the directions of edges if the original graphs are directed and then (2) deleting the nodes that are not contained in the largest connected component of the original graph.
All the following experiments are unaffected because the above pre-processing is performed before setting privacy labels and any processing is not added to the graph after the privacy labels are set. 
The Pokec dataset contains all graph data of the Pokec network, including 552525 real private users (about 33.8\%) and the real privacy labels. 
Table \ref{datasets} lists the size and average degree of five datasets. 
Additionally, we use a dataset \cite{kurant} of 1,016,275 samples of real public Facebook users via random walk during October 2010 to evaluate the actual performance of the proposed estimators. This dataset contains the ID, exact public-degree, and exact degree of each sampled public user.

\begin{table}[t]
\caption{Datasets.}
\label{datasets}
\begin{center}
	\begin{tabular}{l c c c c}\hline
	Network & $n$ & $d_{avg}$ & Privacy label setting  \\ \hline
	YouTube \cite{konect} & 1,134,890 & 5.27 & based on Assumption \ref{assumption_private}  \\
	Pokec \cite{snap} & 1,632,803 & 27.32 & real labels \\
	Orkut \cite{snap} & 3,072,441 & 76.28 & based on Assumption \ref{assumption_private} \\
	Facebook \cite{nr} & 3,097,165 & 15.28 & based on Assumption \ref{assumption_private} \\
	LiveJournal \cite{snap} & 3,997,962 & 17.35 & based on Assumption \ref{assumption_private} \\ \hline
  	\end{tabular}
\end{center}
\end{table}

\begin{figure*}[]
      \begin{minipage}{0.245\hsize}
        \begin{center}
          \includegraphics[scale=\figscale]{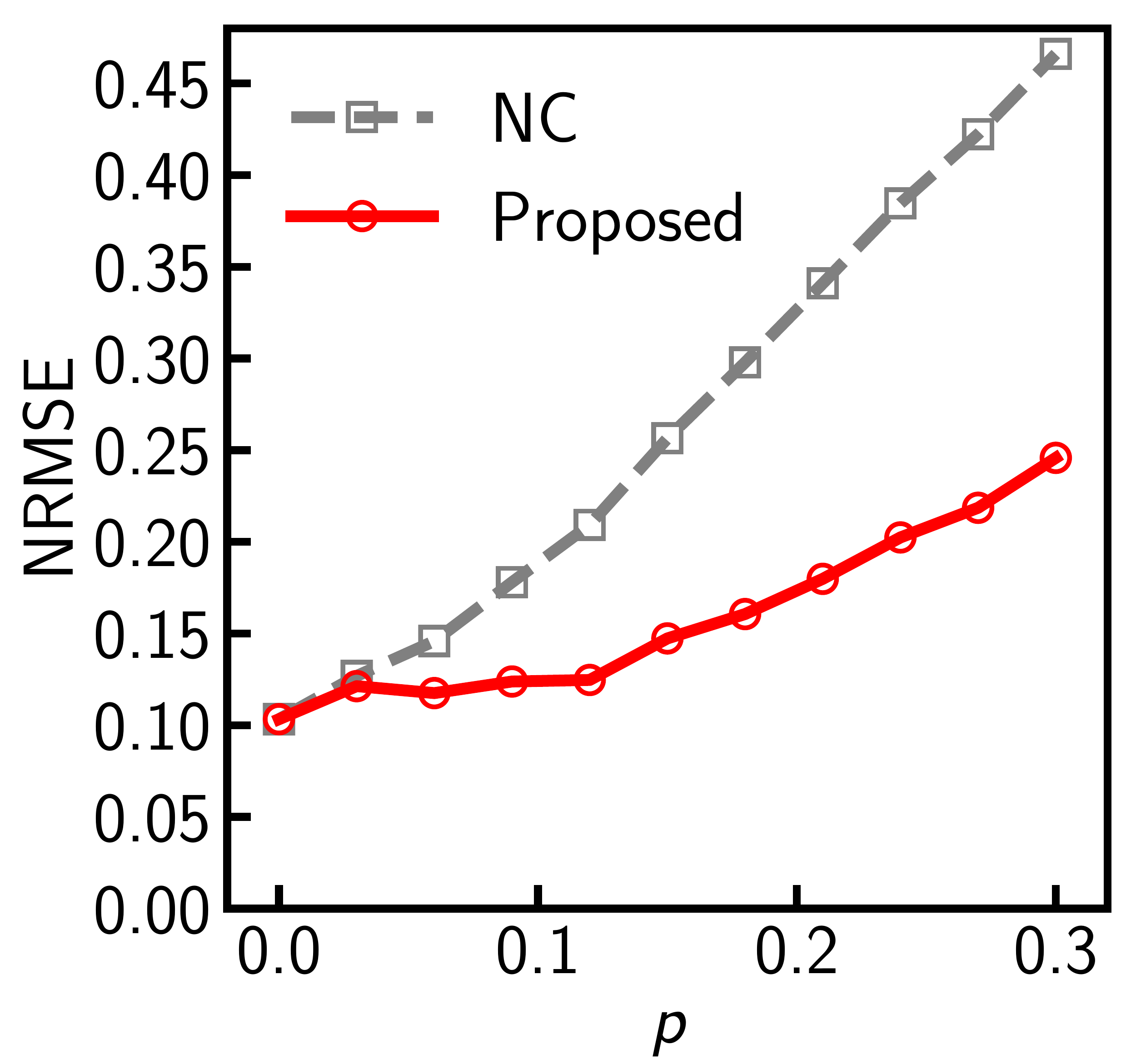}
          \\(a) YouTube
        \end{center}
      \end{minipage}
      \begin{minipage}{0.245\hsize}
        \begin{center}
          \includegraphics[scale=\figscale]{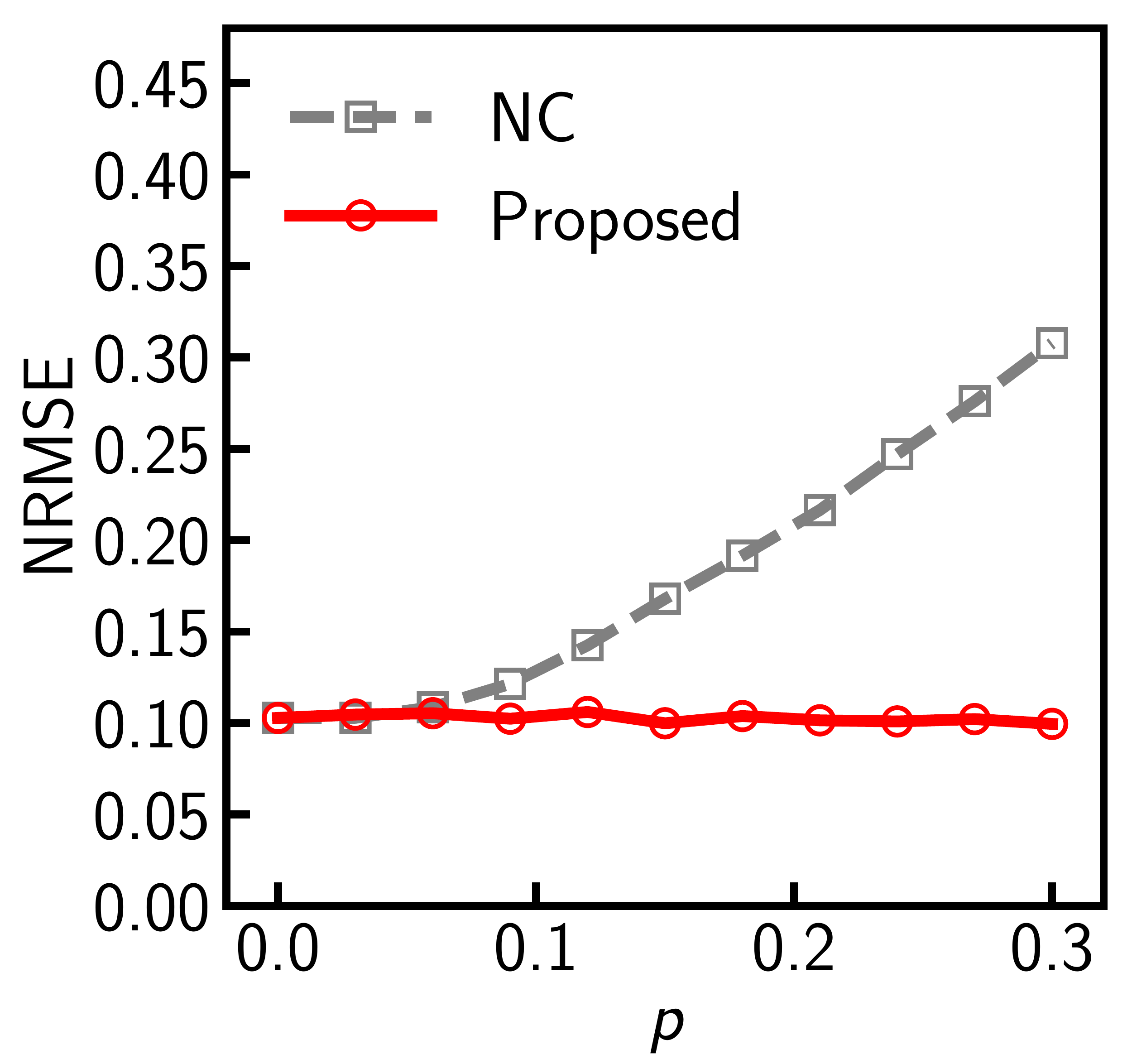}
          \\(b) Orkut
        \end{center}
      \end{minipage}
      \begin{minipage}{0.245\hsize}
        \begin{center}
          \includegraphics[scale=\figscale]{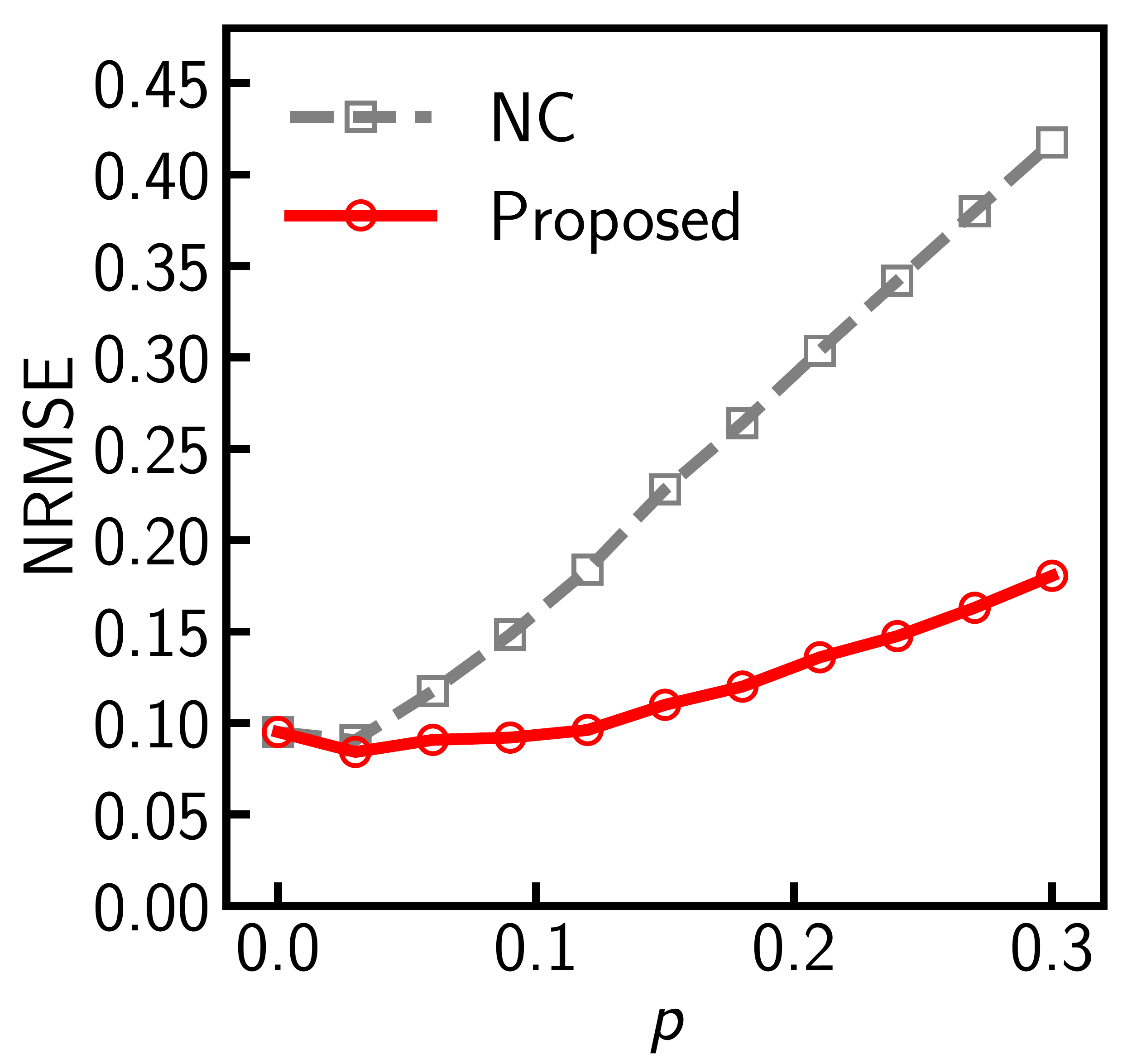}
          \\(c) Facebook
        \end{center}
      \end{minipage}
      \begin{minipage}{0.245\hsize}
        \begin{center}
          \includegraphics[scale=\figscale]{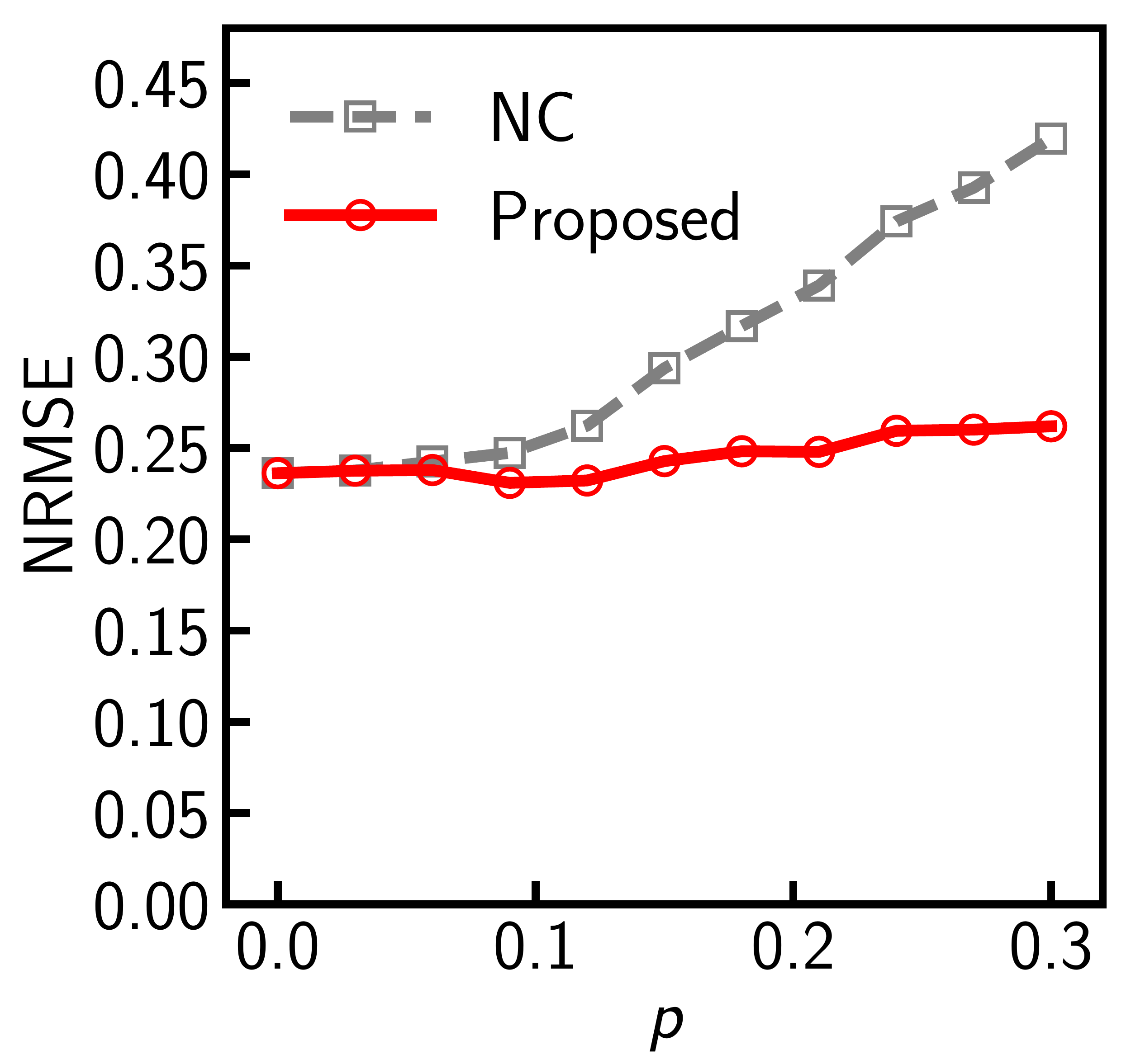}
          \\(d) LiveJournal
        \end{center}
      \end{minipage}
      \vspace{-1mm}
      \caption{NRMSEs of the size estimates (the ideal model with 1\% sample size).}
      \label{size_ideal}
      \vspace{3mm}
	 
      \begin{minipage}{0.245\hsize}
        \begin{center}
          \includegraphics[scale=\figscale]{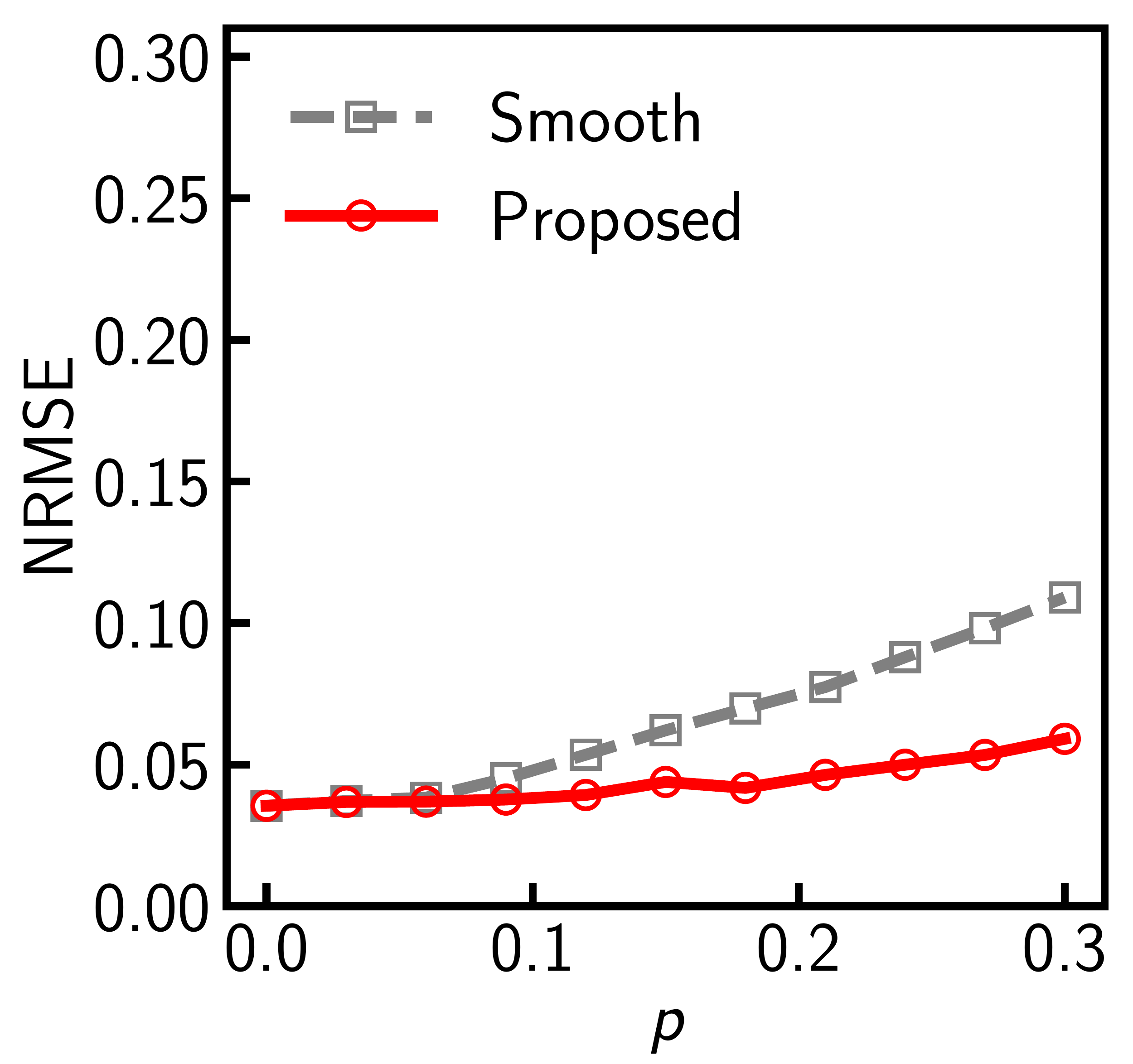}
          \\(a) YouTube
        \end{center}
      \end{minipage}
      \begin{minipage}{0.245\hsize}
        \begin{center}
          \includegraphics[scale=\figscale]{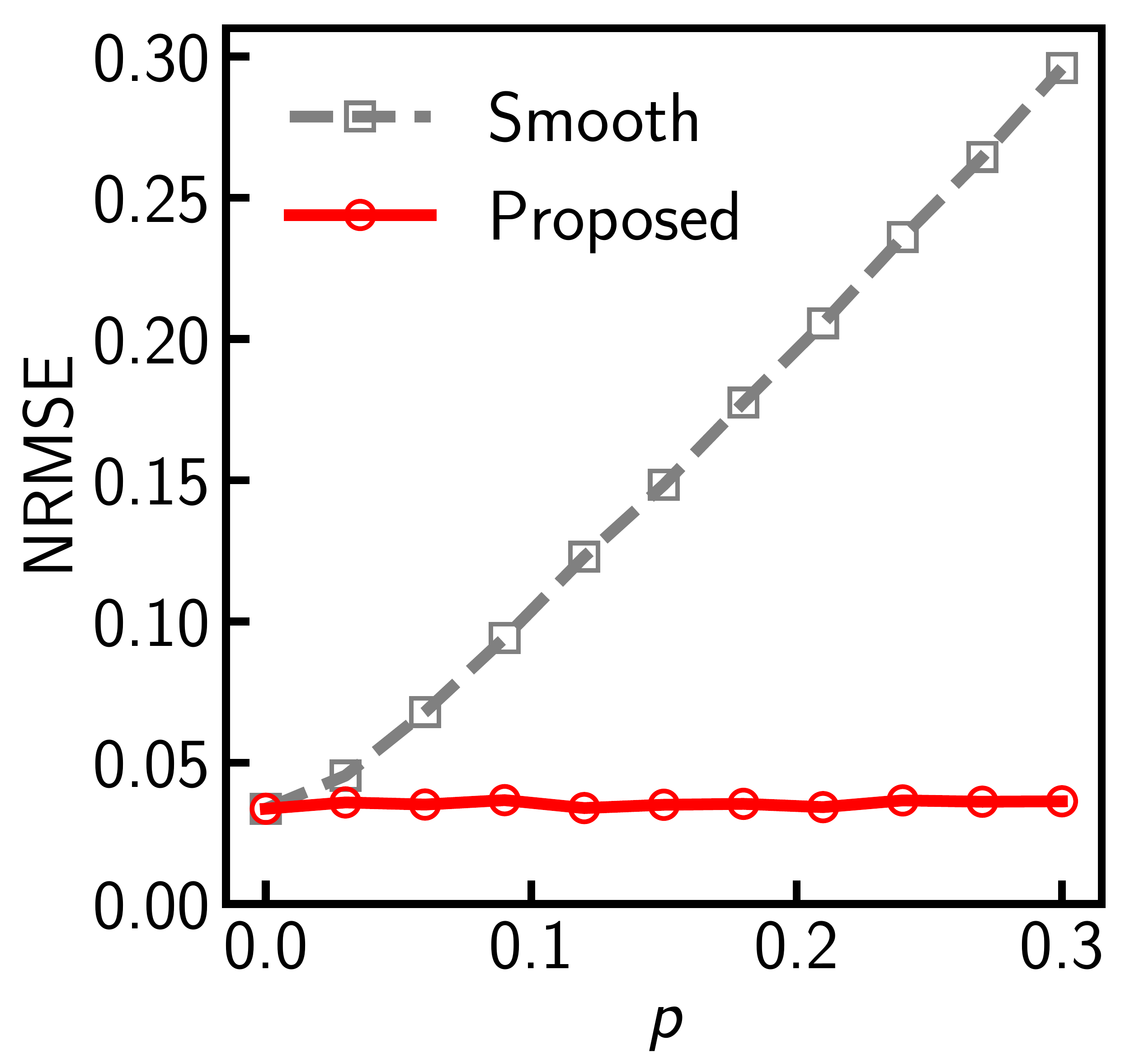}
          \\(b) Orkut
        \end{center}
      \end{minipage}
      \begin{minipage}{0.245\hsize}
        \begin{center}
          \includegraphics[scale=\figscale]{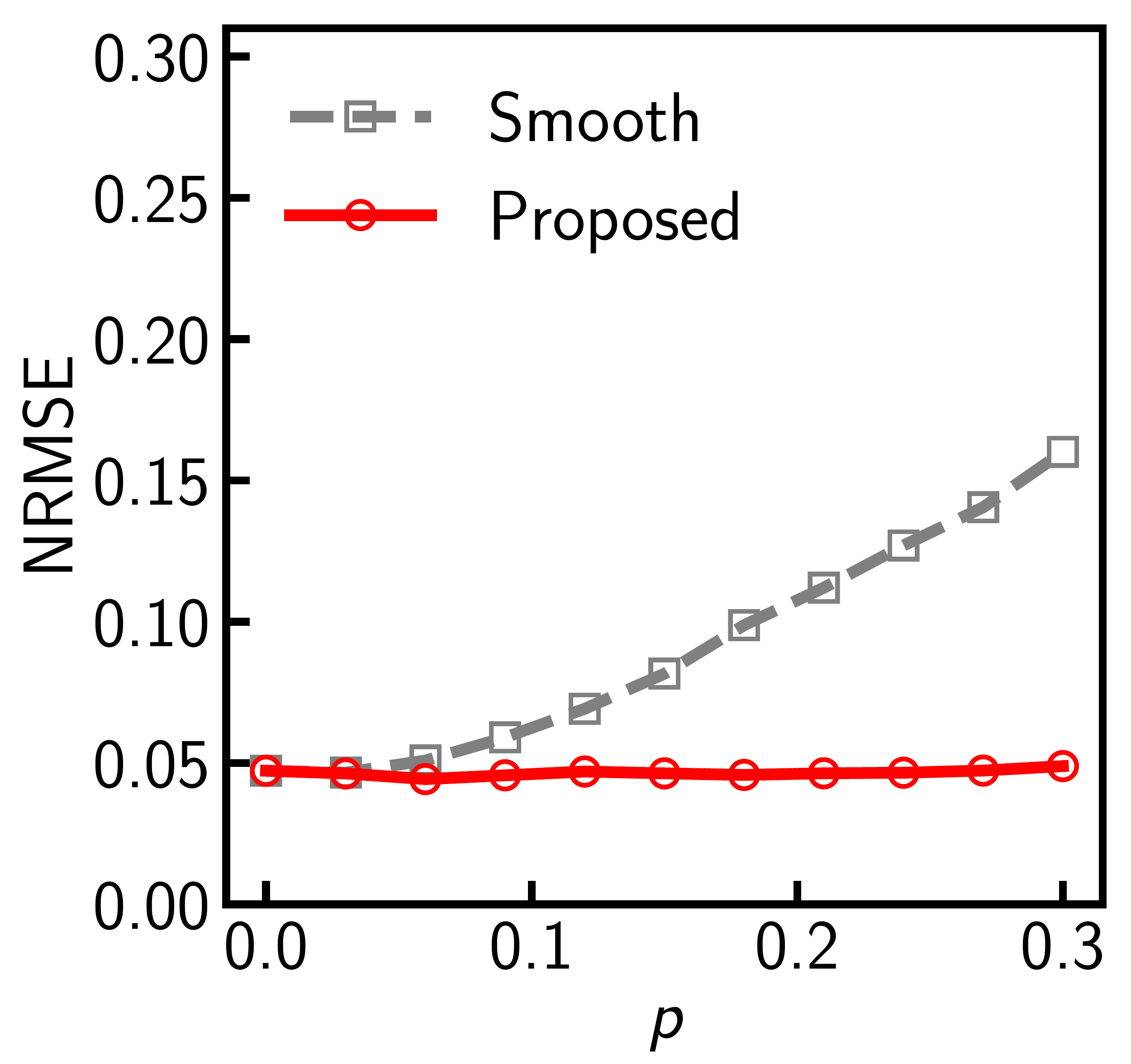}
          \\(c) Facebook
        \end{center}
      \end{minipage}
      \begin{minipage}{0.245\hsize}
        \begin{center}
          \includegraphics[scale=\figscale]{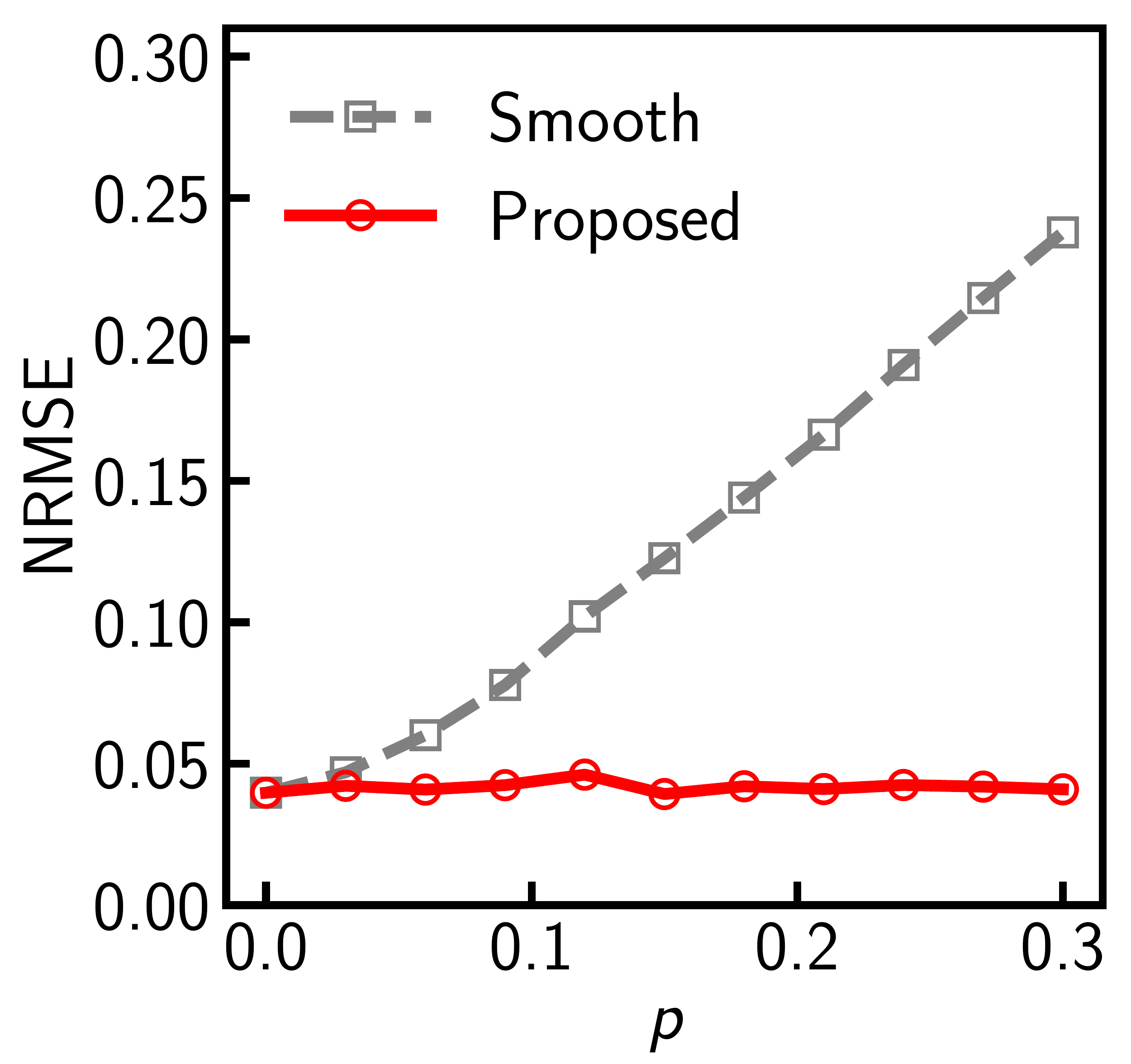}
          \\(d) LiveJournal
        \end{center}
      \end{minipage}
      \vspace{-1mm}
      \caption{NRMSEs of the average degree estimates (the ideal model with 1\% sample size).}
      \label{aved_ideal}
      \vspace{3mm}
      
      \begin{minipage}{0.245\hsize}
        \begin{center}
          \includegraphics[scale=\figscale]{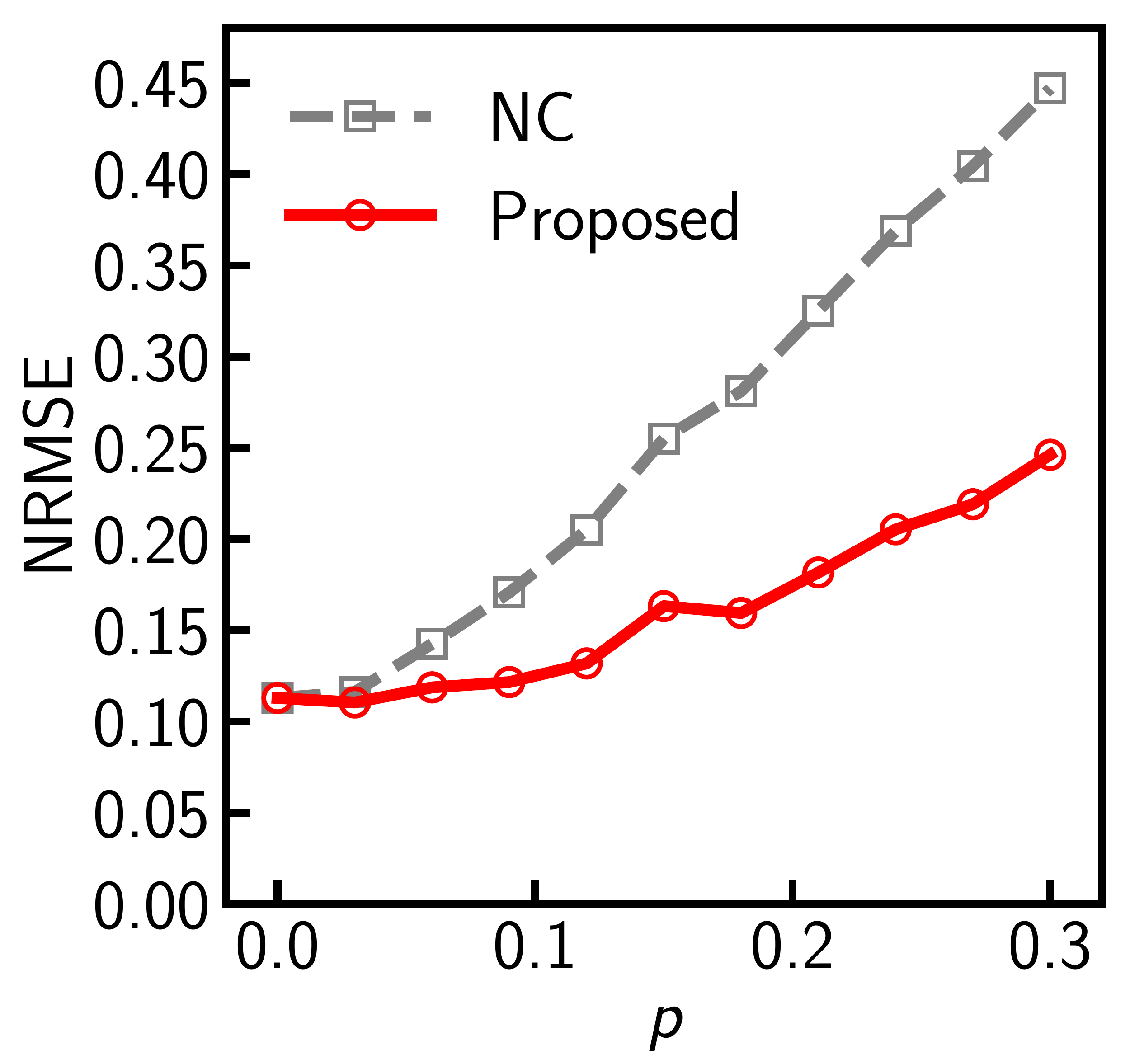}
          \\(a) YouTube
        \end{center}
      \end{minipage}
      \begin{minipage}{0.245\hsize}
        \begin{center}
          \includegraphics[scale=\figscale]{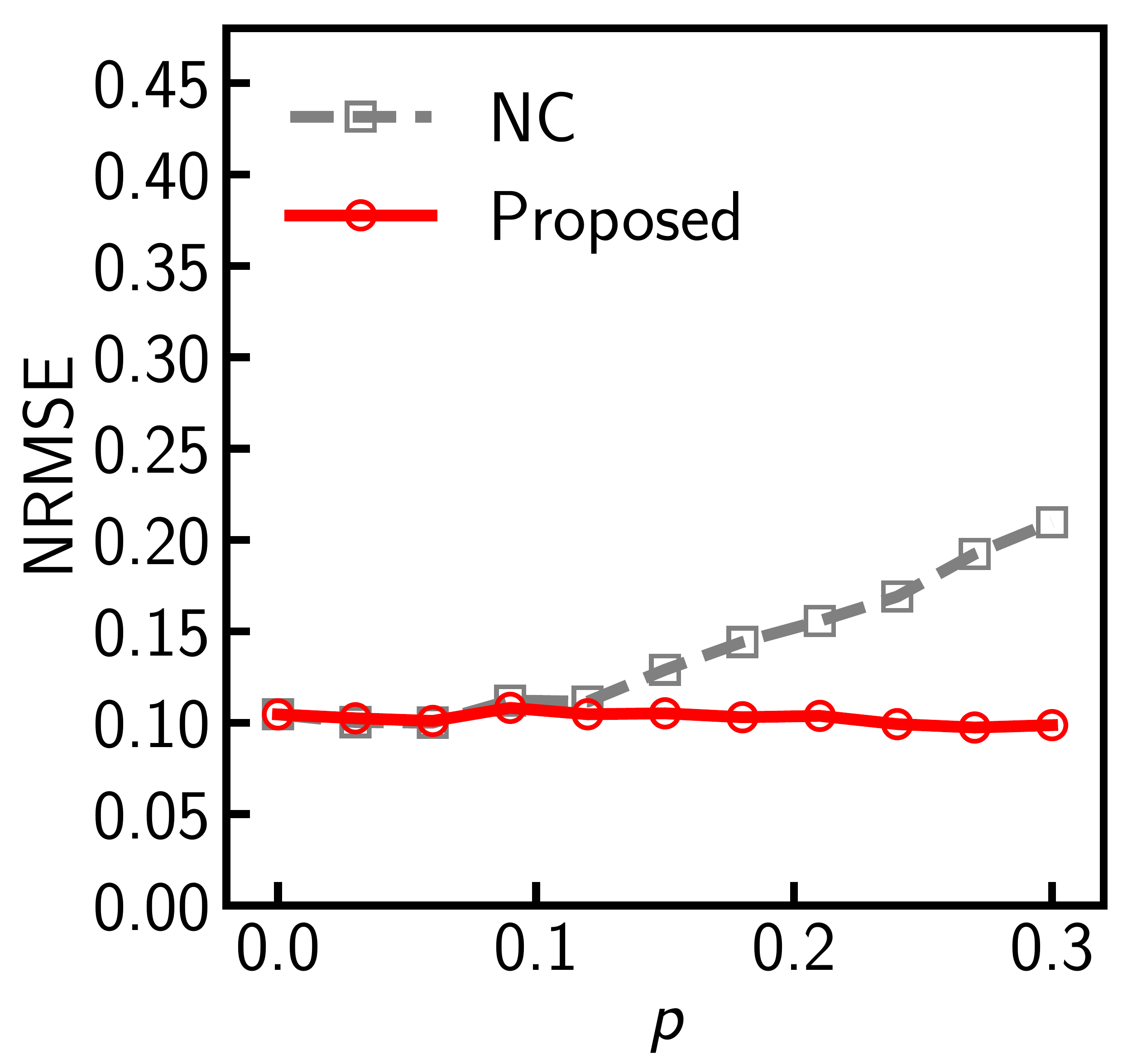}
          \\(b) Orkut
        \end{center}
      \end{minipage}
      \begin{minipage}{0.245\hsize}
        \begin{center}
          \includegraphics[scale=\figscale]{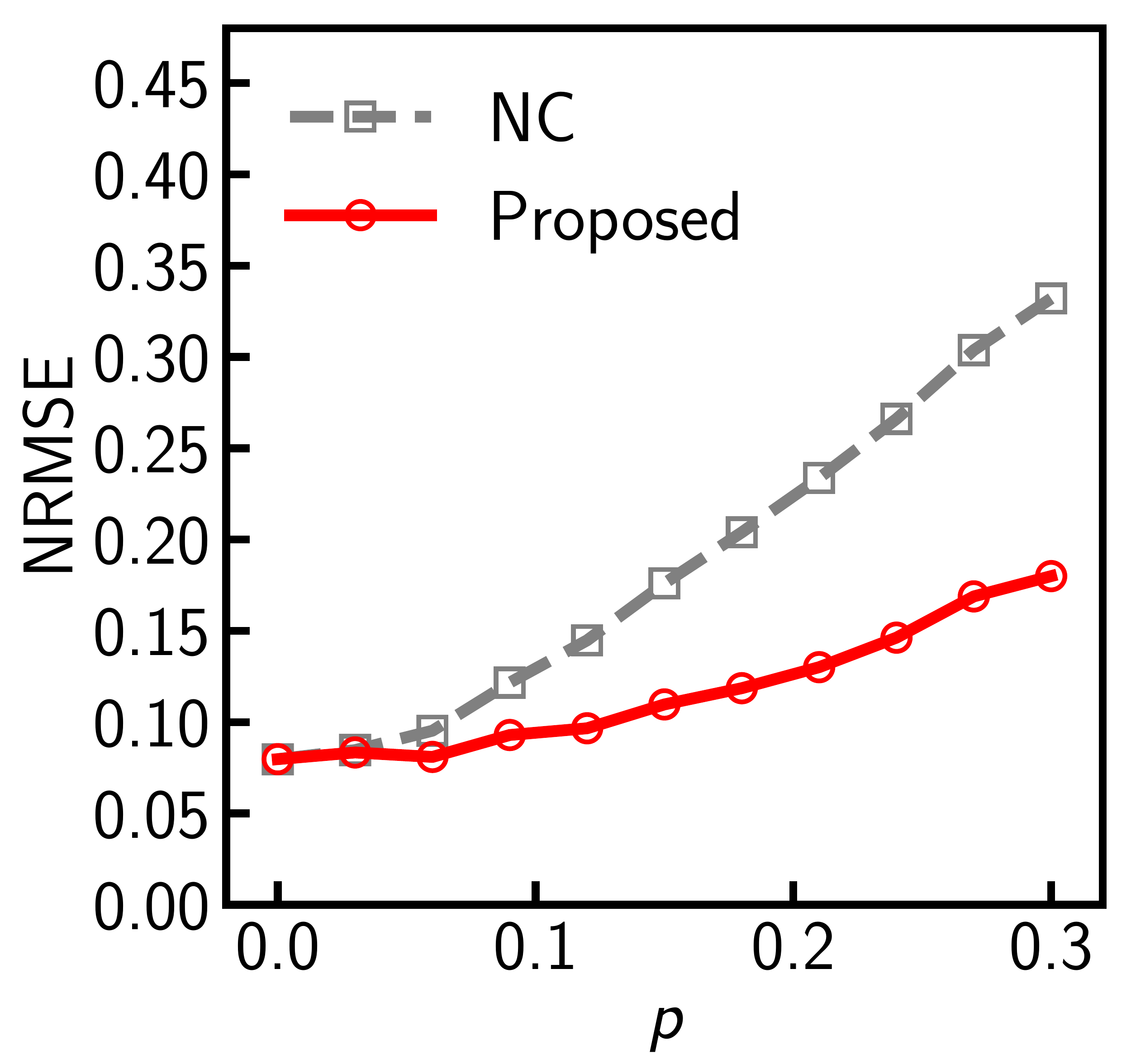}
          \\(c) Facebook
        \end{center}
      \end{minipage}
      \begin{minipage}{0.245\hsize}
        \begin{center}
          \includegraphics[scale=\figscale]{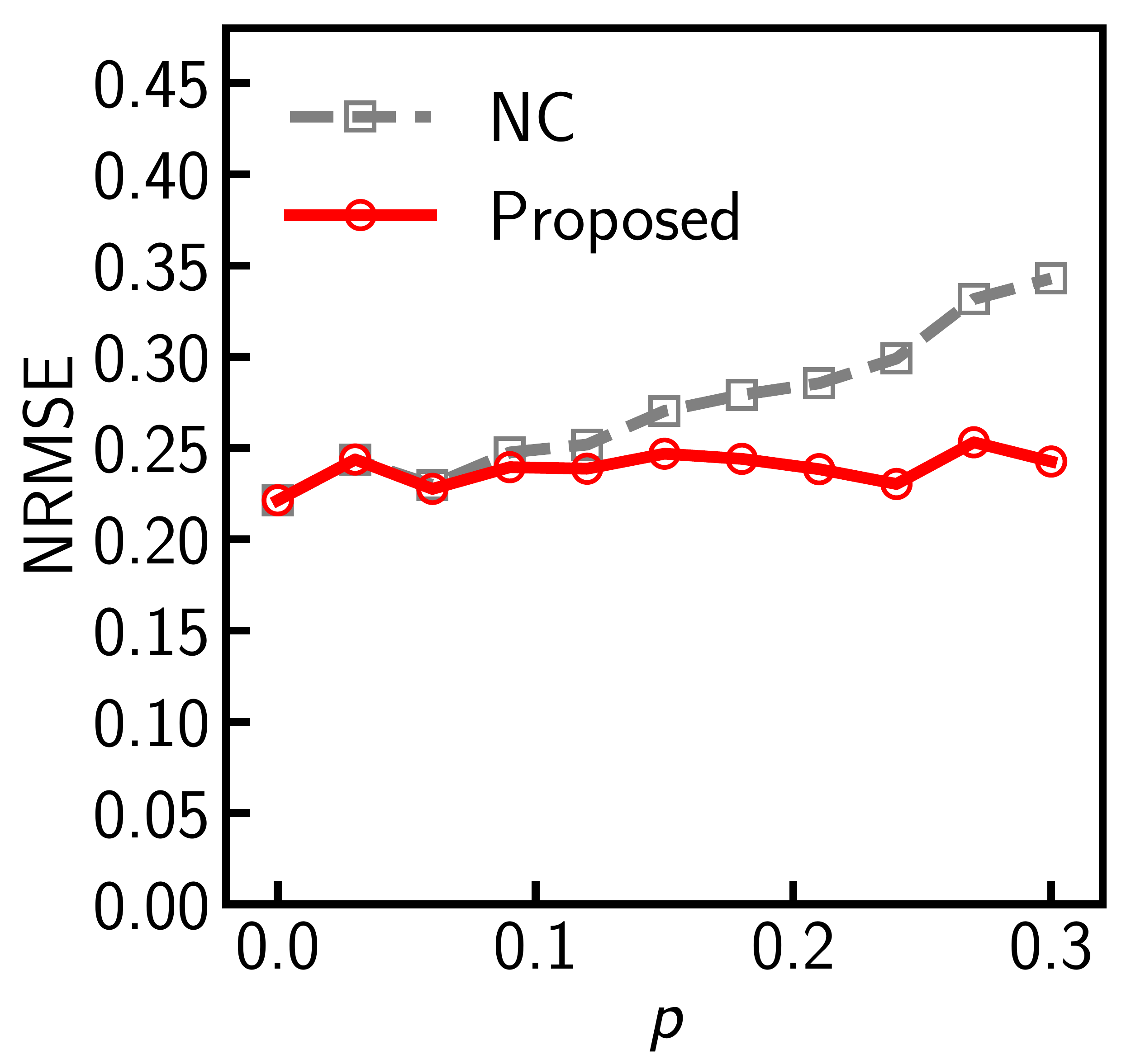}
          \\(d) LiveJournal
        \end{center}
      \end{minipage}
      \vspace{-1mm}
      \caption{NRMSEs of the size estimates (the hidden privacy model with 1\% sample size).}
      \label{size_hidden}
      \vspace{3mm}

      \begin{minipage}{0.245\hsize}
        \begin{center}
          \includegraphics[scale=\figscale]{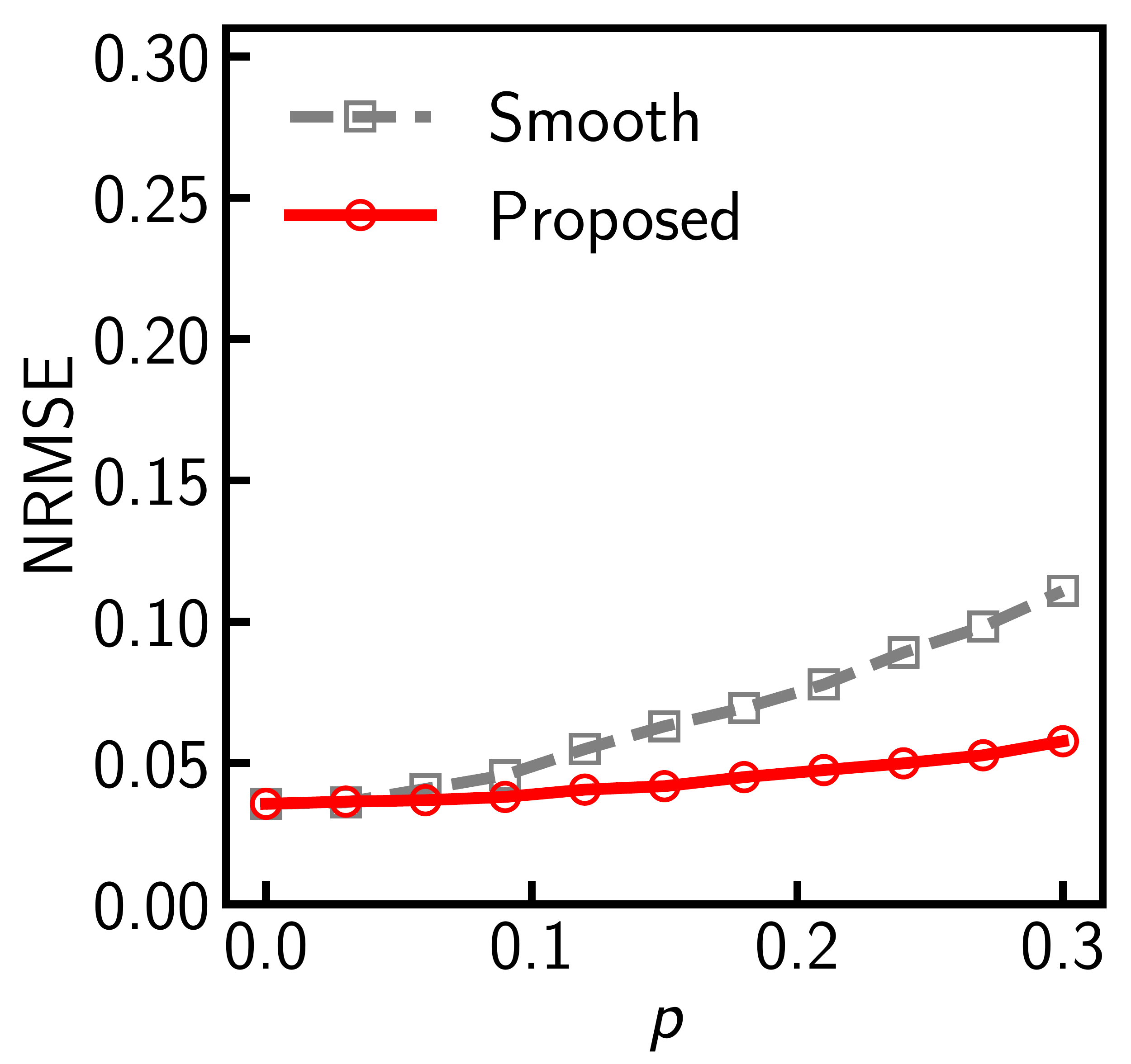}
          \\(a) YouTube
        \end{center}
      \end{minipage}
      \begin{minipage}{0.245\hsize}
        \begin{center}
          \includegraphics[scale=\figscale]{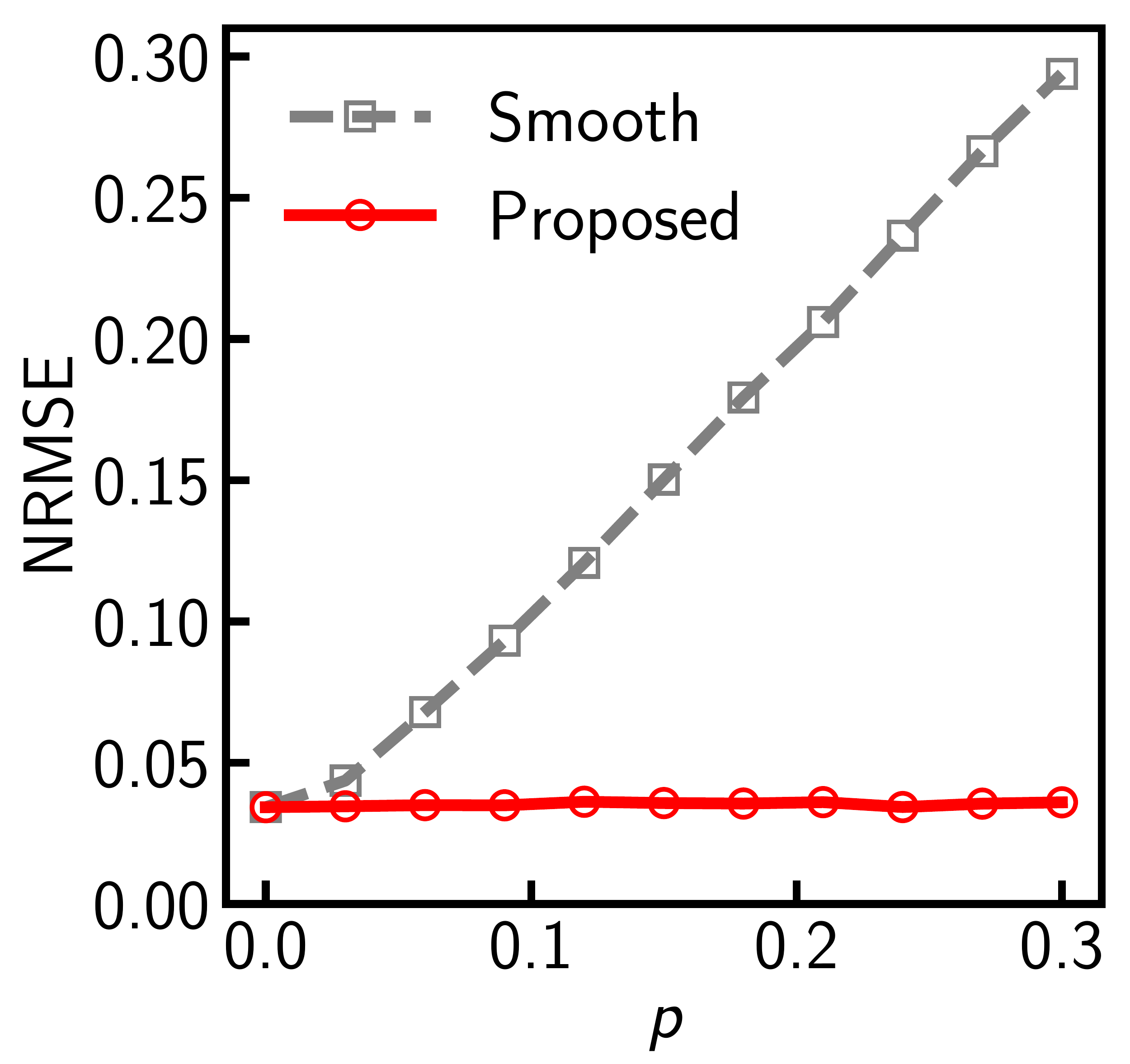}
          \\(b) Orkut
        \end{center}
      \end{minipage}
      \begin{minipage}{0.245\hsize}
        \begin{center}
          \includegraphics[scale=\figscale]{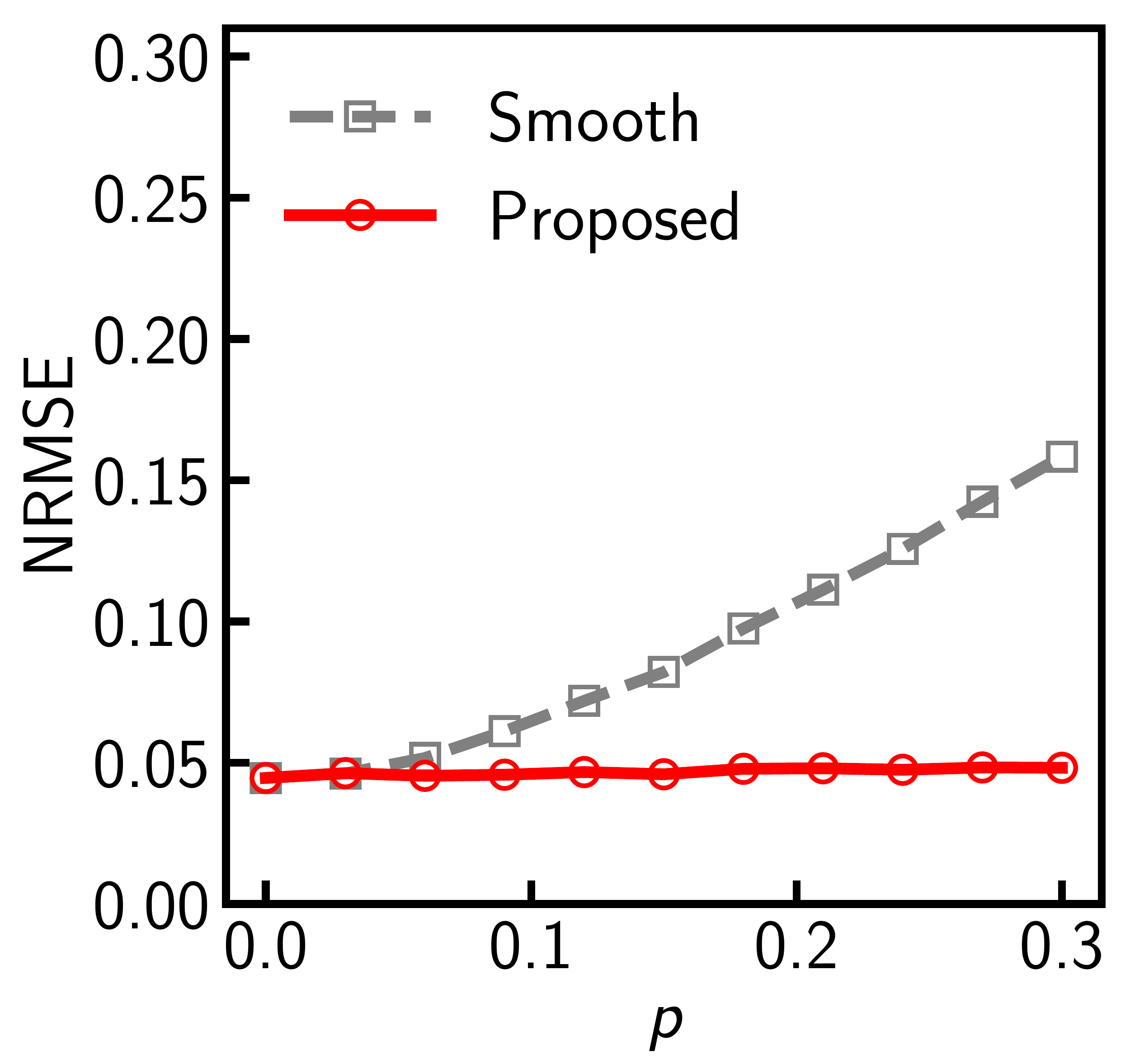}
          \\(c) Facebook
        \end{center}
      \end{minipage}
      \begin{minipage}{0.245\hsize}
        \begin{center}
          \includegraphics[scale=\figscale]{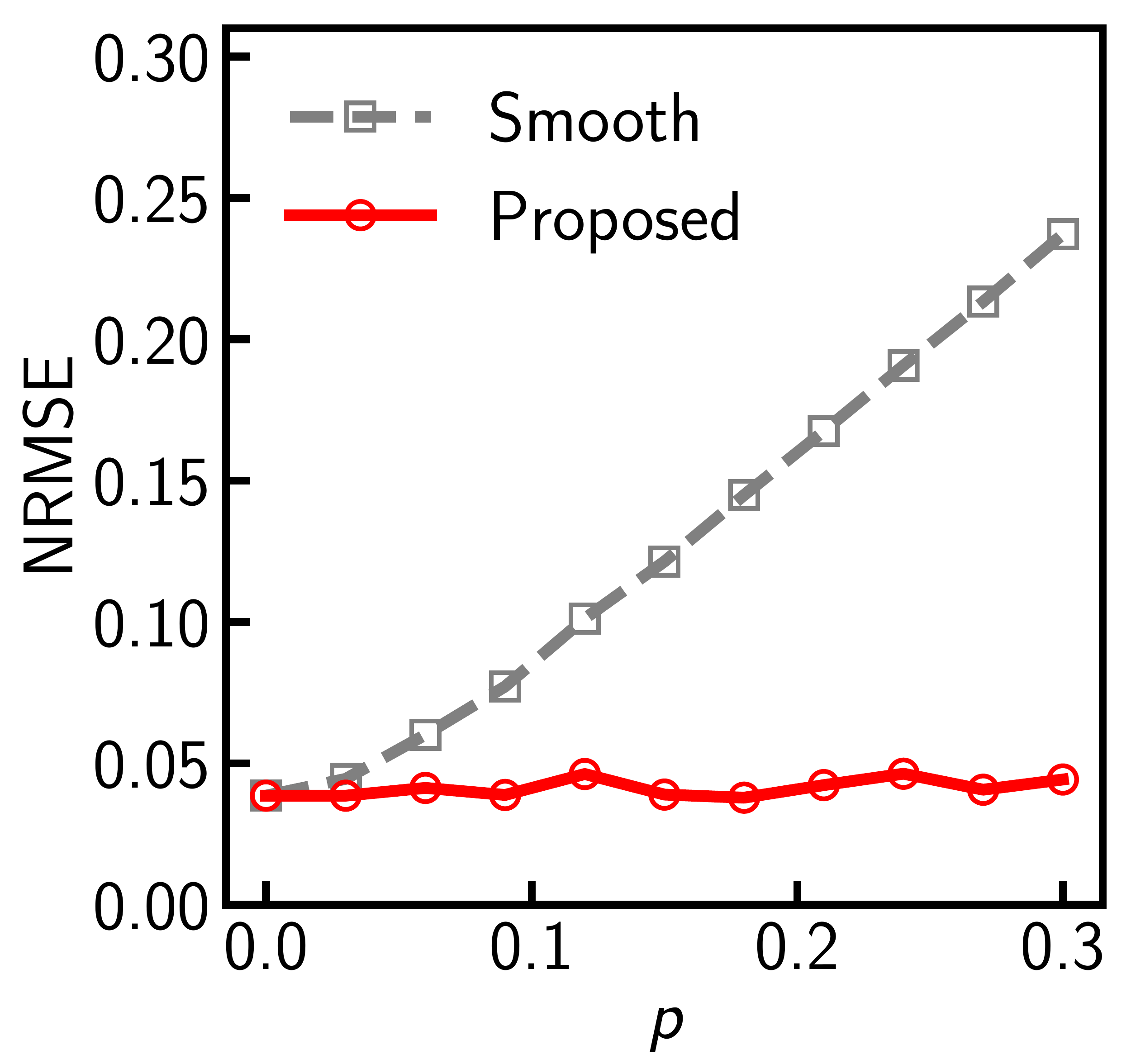}
          \\(d) LiveJournal
        \end{center}
      \end{minipage}
      \vspace{-1mm}
      \caption{NRMSEs of the average degree estimates (the hidden privacy model with 1\% sample size).}
      \label{aved_hidden}
      \vspace{3mm}

	\begin{minipage}{0.495\hsize}
      \begin{minipage}{0.495\hsize}
        \begin{center}
          \includegraphics[scale=\figscale]{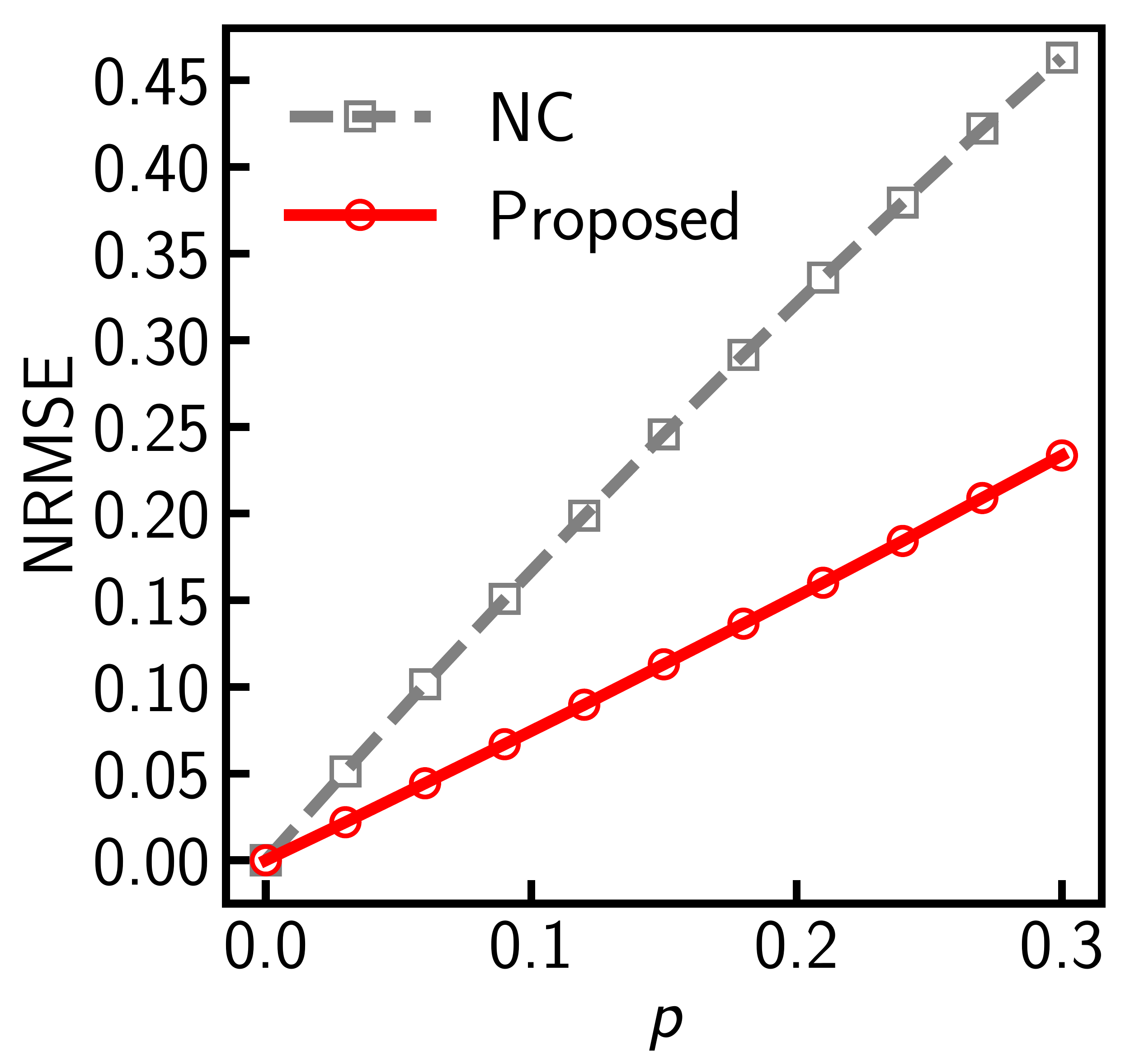}
          \\(a) YouTube
        \end{center}
      \end{minipage}
      \begin{minipage}{0.495\hsize}
        \begin{center}
          \includegraphics[scale=\figscale]{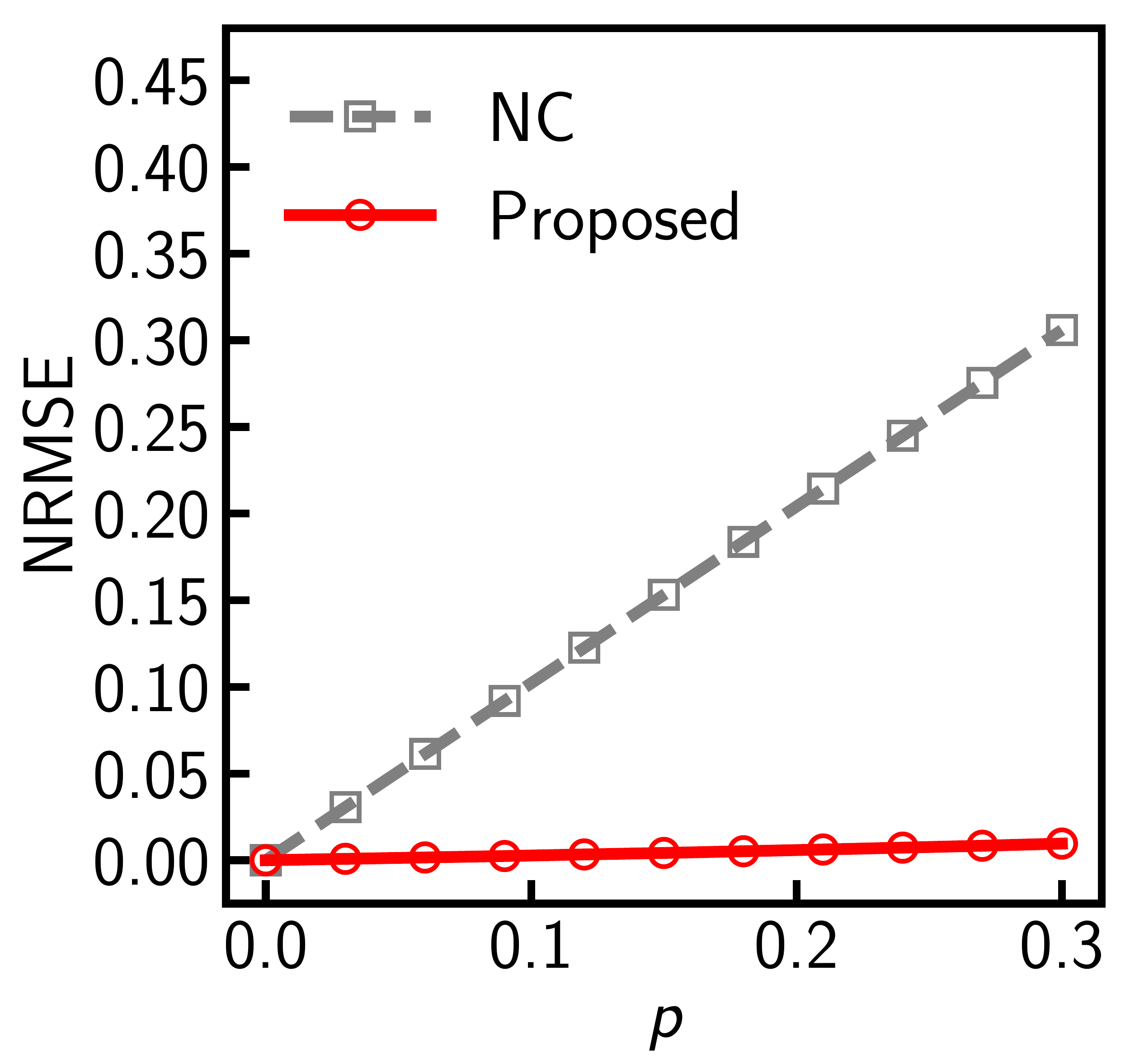}
          \\(b) Orkut
        \end{center}
      \end{minipage}
      \vspace{-1mm}
      \caption{NRMSEs of size convergence values.}
      \label{size_conv}
	\end{minipage}
	\begin{minipage}{0.495\hsize}
      \begin{minipage}{0.495\hsize}
        \begin{center}
          \includegraphics[scale=\figscale]{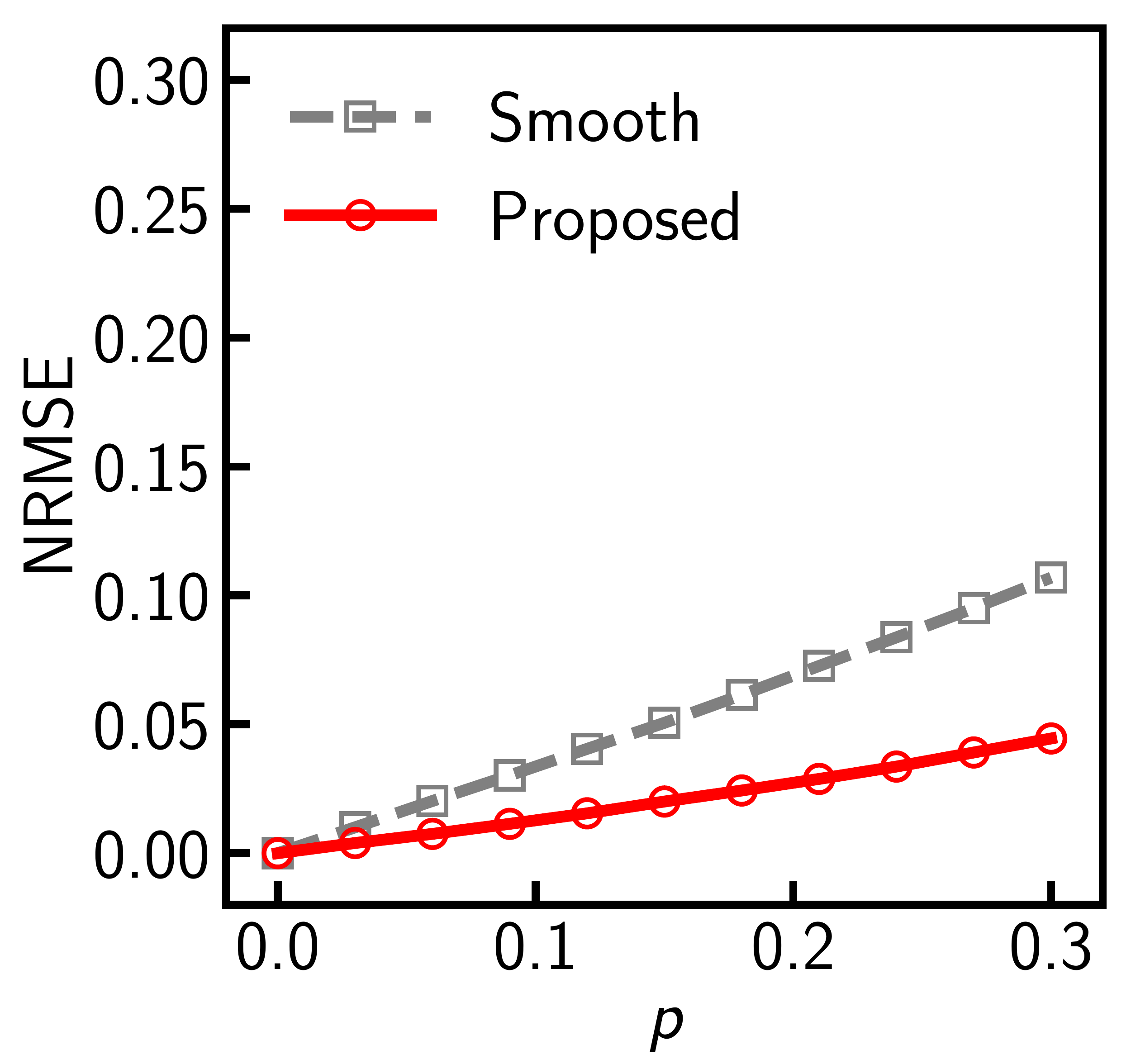}
          \\(a) YouTube
        \end{center}
      \end{minipage}
      \begin{minipage}{0.495\hsize}
        \begin{center}
          \includegraphics[scale=\figscale]{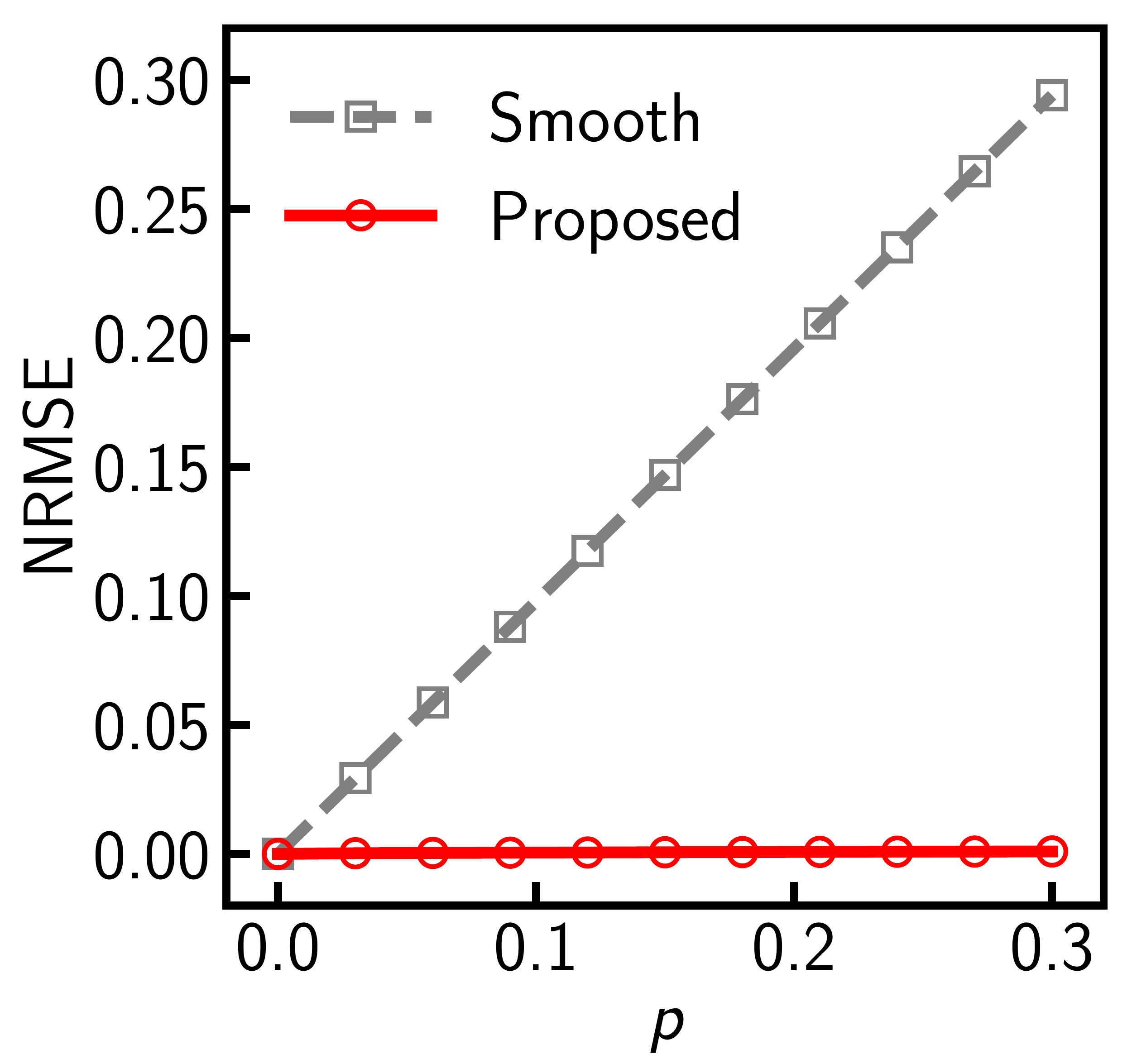}
          \\(b) Orkut
        \end{center}
      \end{minipage}
      \vspace{-1mm}
      \caption{NRMSEs of average degree convergence values.}
      \label{aved_conv}
	\end{minipage}
\end{figure*}

{\bf Accuracy measure:} We apply the normalized root mean square error (NRMSE) to evaluate both the bias and variance of estimators \cite{chen, katzir_nodecc, lee_nonback, wang}. NRMSE is defined by $(E[(\frac{\tilde{x}}{x} - 1)^2])^{\frac{1}{2}}$, where $\tilde{x}$ denotes the estimation value and $x$ denotes the true value. We performed each experiment independently 1000 times to estimate the NRMSE.

{\bf Privacy label settings:} For YouTube, Orkut, Facebook, and LiveJournal, we independently set each privacy label as private with probability $p$ and otherwise public, according to Assumption \ref{assumption_private}. The probability $p$ is from 0.0 to 0.30 in increments of 0.03 because there were actually tens of percentages of private users \cite{catanese, dey, takac}. The set of privacy labels is independently given for each experiment. For Pokec, we apply the set of real privacy labels contained in the dataset. Table \ref{datasets} lists how to set privacy labels in five datasets.

{\bf Seed selection:} On YouTube, Orkut, Facebook, LiveJournal, and Pokec, the seed of a random walk is selected randomly from nodes on the largest public-cluster and independently for each experiment.

{\bf Sample size: } The sample size, i.e., the length of a random walk, $r$, is 1\% of the total number of nodes on YouTube, Orkut, Facebook, and LiveJournal. On Pokec, the sample size is varied from 0.5\% to 5\% of the total number of nodes in increments of 0.5\%.

{\bf Threshold in size estimators:} The threshold, $m$, in size estimators is 2.5\% of the sample size, as set in a previous study \cite{katzir_nodecc}.

{\bf Degree distributions:} Figure \ref{degree_distribution} shows the cumulative degree distributions of five datasets. We see that the degree distribution is biased to low degrees in all datasets.

{\bf Coefficients $\alpha_p$:} Figure \ref{alpha_p} shows the coefficients $\alpha_p$ that is defined in Theorem \ref{error_size_ours} for various values of $p$ in five datasets. We see that the coefficients $\alpha_p$ are almost equal to 1.0 for every value of $p$.

{\bf Relative size of the largest public-cluster:} Figure \ref{cluster_size} shows the average relative size of the largest public-cluster, $\frac{n^*}{n}$, over independent 1000 experiments for various probabilities of $p$ in four datasets. The upper limit is $1 - p$ (gray solid line). All the public nodes on Pokec belong to the largest public-cluster.

{\bf Average absolute size of isolated public-clusters:} Figure \ref{isolated_size} shows the average absolute size of isolated public-clusters (i.e., public-clusters other than the largest public-cluster) over independent 1000 experiments for various probabilities of $p$ in four datasets. Pokec has no isolated public-clusters.

\subsection{Estimation Accuracy for Percentage of Private Nodes} \label{performance}
Figures \ref{size_ideal} and \ref{aved_ideal} show the NRMSEs of each estimator for various probabilities of $p$ with 1\% sample size in the ideal model. 
The NRMSEs of both estimators are equal when $p = 0$ because of Propositions \ref{p0_size} and \ref{p0_aved}.
The proposed estimators have lower NRMSEs for most probabilities of $p>0$ than those of the existing estimators. 
For example, the proposed estimators improve the NRMSE approximately 88.1\%, e.g., from 0.294 to 0.035, for $p = 0.3$ in Figure \ref{aved_ideal} (b).
Figures \ref{size_hidden} and \ref{aved_hidden} show that the proposed estimators improve the NRMSEs on all four datasets in the hidden privacy model as well. 

The improvement of estimation errors results from that of expected errors of convergence values.
Figures \ref{size_conv} and \ref{aved_conv} show the NRMSEs of convergence values obtained from Lemmas \ref{the_size_exp_naive}, \ref{the_size_exp_ours}, \ref{the_aved_exp_naive} and \ref{the_aved_exp_ours} for various probabilities of $p$ on YouTube and Orkut. The NRMSEs of both estimators are equal to 0 when $p = 0$ because of Propositions \ref{p0_size} and \ref{p0_aved}.
The proposed estimators improve the NRMSEs of convergence values for any $p > 0$; this supports the claims of Corollaries \ref{relative_size} and \ref{relative_aved}. We have obtained similar results on Facebook and LiveJournal.

The errors of estimation and convergence values are affected by the relative size of the largest public-cluster because we sample the nodes only on the largest public-cluster. 
On Orkut where almost all the public nodes belong to the largest public-cluster (see Figure \ref{cluster_size}), the convergence values of proposed estimators are almost equal to true values for any probabilities of $p$ (see Figures \ref{size_conv} and \ref{aved_conv}); this supports the claims of Theorems \ref{error_size_ours} and \ref{error_aved_ours}.
Consequently, the NRMSEs of estimates on Orkut do not almost increase as the probabilities of $p$ increases (see Figures \ref{size_ideal}, \ref{aved_ideal}, \ref{size_hidden}, and \ref{aved_hidden}).
Conversely, on YouTube where there are the most public nodes that do not belong to the largest public-cluster among four datasets, the NRMSEs of the estimation and convergence values relatively increases.

\subsection{Estimation in Real-World Datasets} \label{realdataset}
We evaluate the performance of the proposed estimators in two realistic datasets including real private users.

\subsubsection{Pokec} \label{pokec}
We use the Pokec dataset \cite{snap, takac} that contains all the graph data and real privacy labels of the Pokec network. 
There are 552525 private nodes (approximately 33.8\%) on the Pokec graph.
We evaluate errors of estimates and convergence values obtained by the existing and proposed estimators using this dataset. 

\begin{figure}[t]
	  \begin{minipage}{0.495\hsize}
        \begin{center}
          \includegraphics[scale=\figscale]{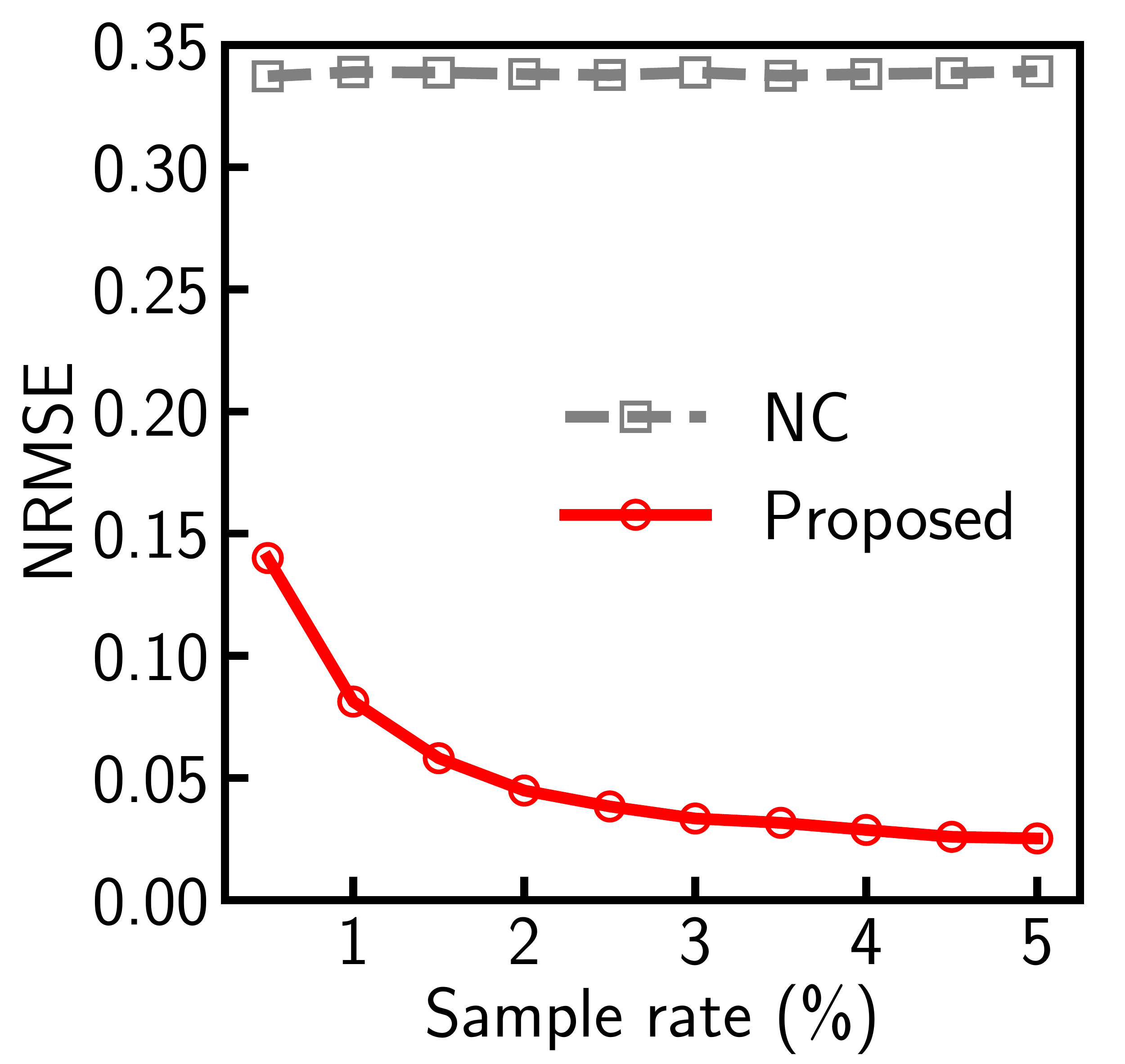}
          \\(a) The ideal model
        \end{center}
      \end{minipage}
      \begin{minipage}{0.495\hsize}
        \begin{center}
          \includegraphics[scale=\figscale]{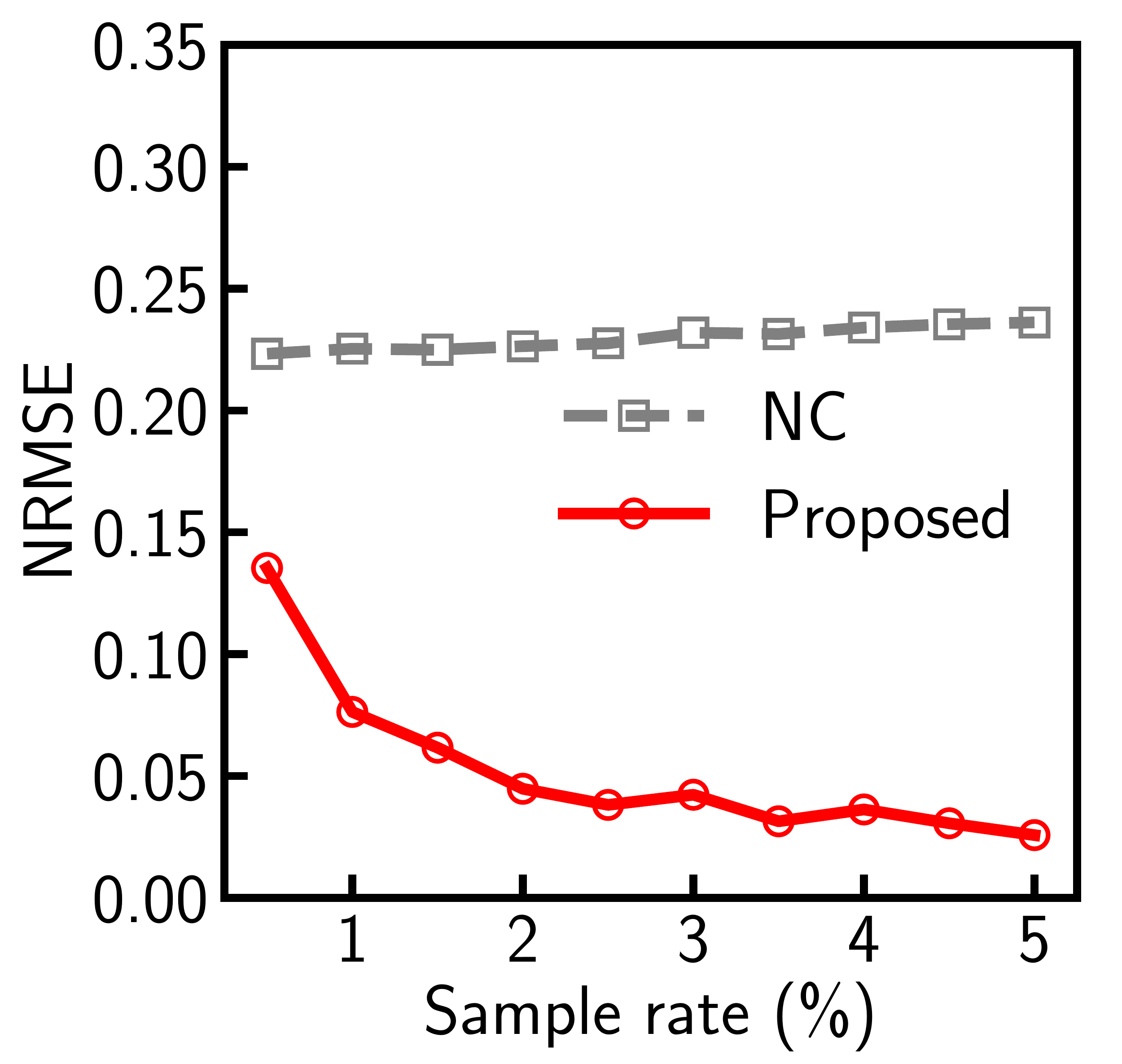}
          \\(b) The hidden privacy model
        \end{center}
      \end{minipage}
      \caption{NRMSEs of size estimates on Pokec.}
      \label{pokec_size}
\end{figure}

\begin{figure}[t]
      \begin{minipage}{0.495\hsize}
        \begin{center}
          \includegraphics[scale=\figscale]{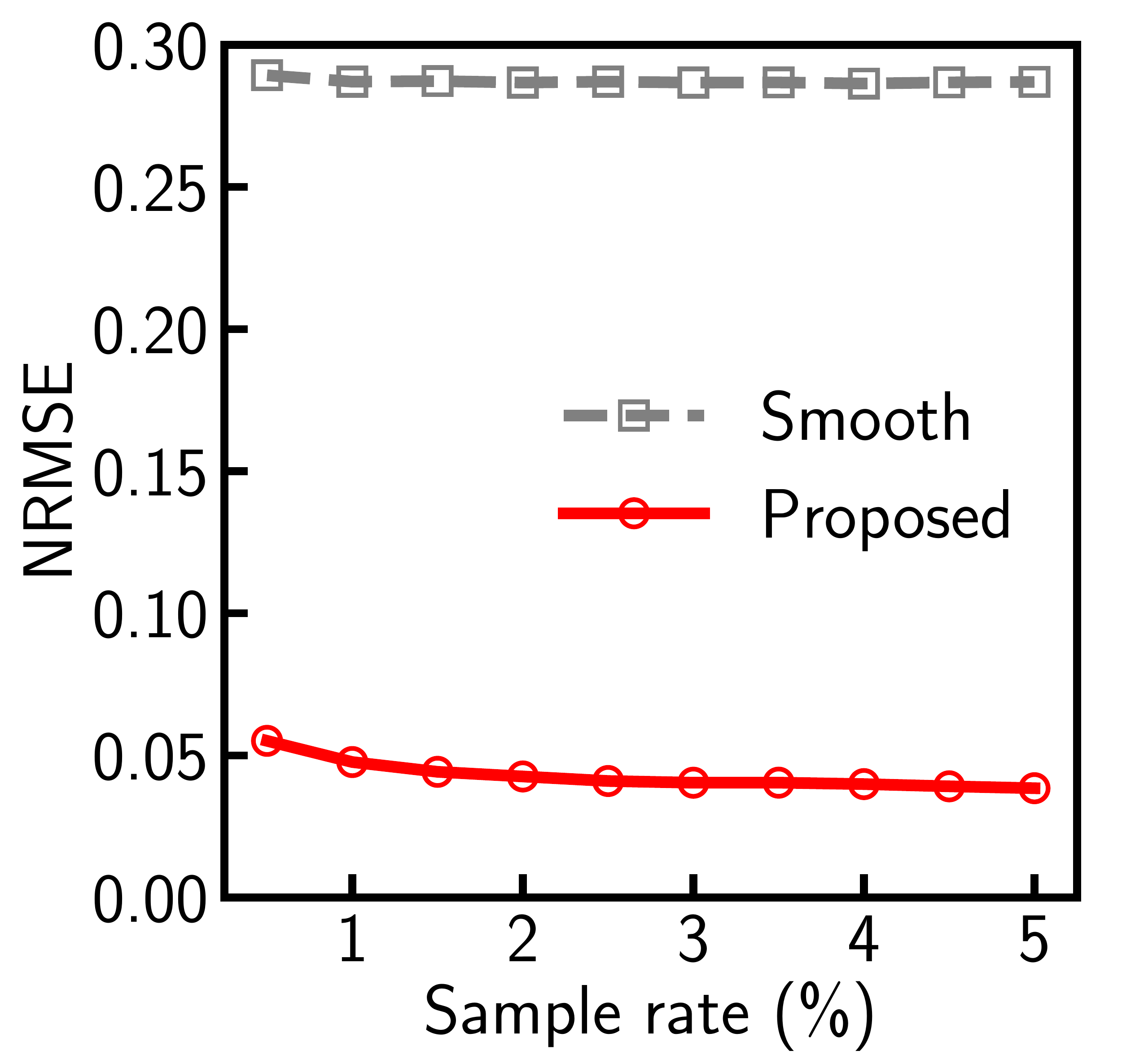}
          \\(a) The ideal model
        \end{center}
      \end{minipage}
      \begin{minipage}{0.495\hsize}
        \begin{center}
          \includegraphics[scale=\figscale]{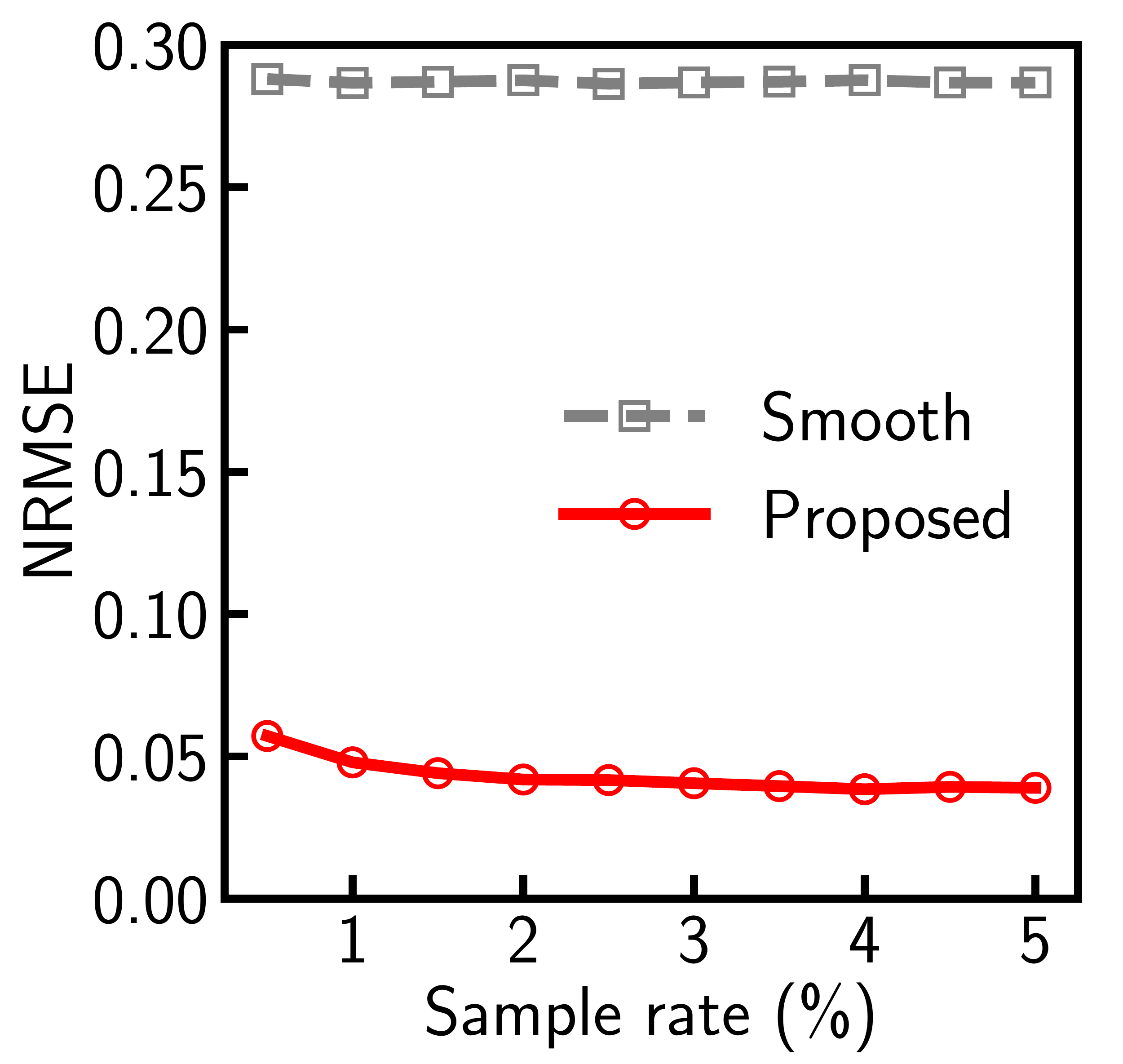}
          \\(b) The hidden privacy model
        \end{center}
      \end{minipage}
      \caption{NRMSEs of average degree estimates on Pokec.}
      \label{pokec_aved}
\end{figure}

Figures \ref{pokec_size} and \ref{pokec_aved} show the NRMSEs of each estimator for  various sample sizes.
In both access models, the proposed estimators improve the NRMSEs for all sample sizes. 
The proposed size estimator particularly improves the NRMSE by approximately 92.6\%, e.g., from 0.339 to 0.025, with a 5\% sample size in Figure \ref{pokec_size} (a). 

The improvement of estimation errors results from that of the errors of convergence values. 
The left column in Table \ref{real_result} presents the convergence values obtained from Lemmas \ref{the_size_exp_naive}, \ref{the_size_exp_ours}, \ref{the_aved_exp_naive} and \ref{the_aved_exp_ours} and relative errors of each estimator. 
The proposed estimators improve the relative errors by 97.3\%, e.g., from 0.338 to 0.009, for the size and 87.5\%, e.g., from 0.287 to 0.036 for the average degree. 

Moreover, the proposed estimators have convergence values that are almost equal to the true properties of the whole Pokec graph including private nodes.
Even though Assumption \ref{assumption_private} does not hold for the Pokec graph, we have obtained reasonable results as claimed in Theorems \ref{error_size_ours} and \ref{error_aved_ours}.


\begin{table}[t]
\begin{center}
\caption{Convergence values (CV) and relative errors (RE) in Pokec (left column), and  estimates obtained from samples of public Facebook Users (right column).}
\label{real_result}
\begin{tabular}{c | c c | c} \hline
	& \multicolumn{2}{c|}{Pokec} & Real Facebook \\ \hline
	Algorithm & CV & RE & Estimate \\ \hline
	{\bf NC} & 1080278.0 & 0.338 & 480298540.0 \\ 
	{\bf Proposed} & 1646880.8 & {\bf 0.009} & 656874080.9 \\ \hline
	{\bf Smooth} & 19.49 & 0.287 & 102.07 \\
	{\bf Proposed} & 28.31 & {\bf 0.036} & 137.03 \\ \hline
  \end{tabular}
\end{center}
\end{table}

\subsubsection{Real public Facebook user samples} \label{facebook}
We use a dataset \cite{kurant} of 1,016,275 real public Facebook user samples obtained by a random walk during October 2010. 
We can estimate the size and average degree of Facebook as of October 2010 using existing and proposed estimators, because this dataset contains the ID, exact public-degree, and exact degree of each sampled public user.

The right column in Table \ref{real_result} shows estimates of each estimator. 
We set a threshold, $m$, in size estimators as 2.5\% of the sample size, i.e., $m = 25407$. 
Although we cannot evaluate each error because the true size and average degree of Facebook as of October 2010 are unknown, we believe that estimates are reasonable, considering two findings in almost the same period.

First, Facebook reported that there were 500 million {\it active} users as of July 2010 \cite{facebook_500m}. 
This suggests that there were at least 500 million users including inactive users at that time. 
Notably, the estimated size in Table \ref{real_result} includes both active and inactive users. 
Our estimate, i.e., 657 million, is greater than 500 million, and we speculate that the difference (approximately 157 million) includes inactive users.

Second, Catanese et al. obtained an unbiased estimate of the proportion of private users as 0.266 from uniform samples of Facebook users in August 2010 \cite{catanese}. 
From Lemmas \ref{error_size_naive} and \ref{error_aved_naive} and Theorems \ref{error_size_ours} and \ref{error_aved_ours}, we can intuitively estimate the value of $p$, which is almost equal to the proportion of private users, from the respective estimates as $\hat{p}_n \triangleq 1 - \frac{n^{NC}}{\hat{n}}$ and $\hat{p}_{avg} \triangleq 1 - \frac{d_{avg}^{Smooth}}{\hat{d}_{avg}}$. Here, $\hat{p}_n$ and $\hat{p}_{avg}$ are the estimates of $p$ obtained by the existing and proposed estimators for the size and average degree, respectively. From Table \ref{real_result}, we obtain $\hat{p}_n$ and $\hat{p}_{avg}$ as 0.269 and 0.255, respectively, which are considerably close to the ground truth value of 0.266.

\begin{figure}[t]
      \begin{minipage}{0.48\hsize}
        \begin{center}
          \includegraphics[scale=\figscale]{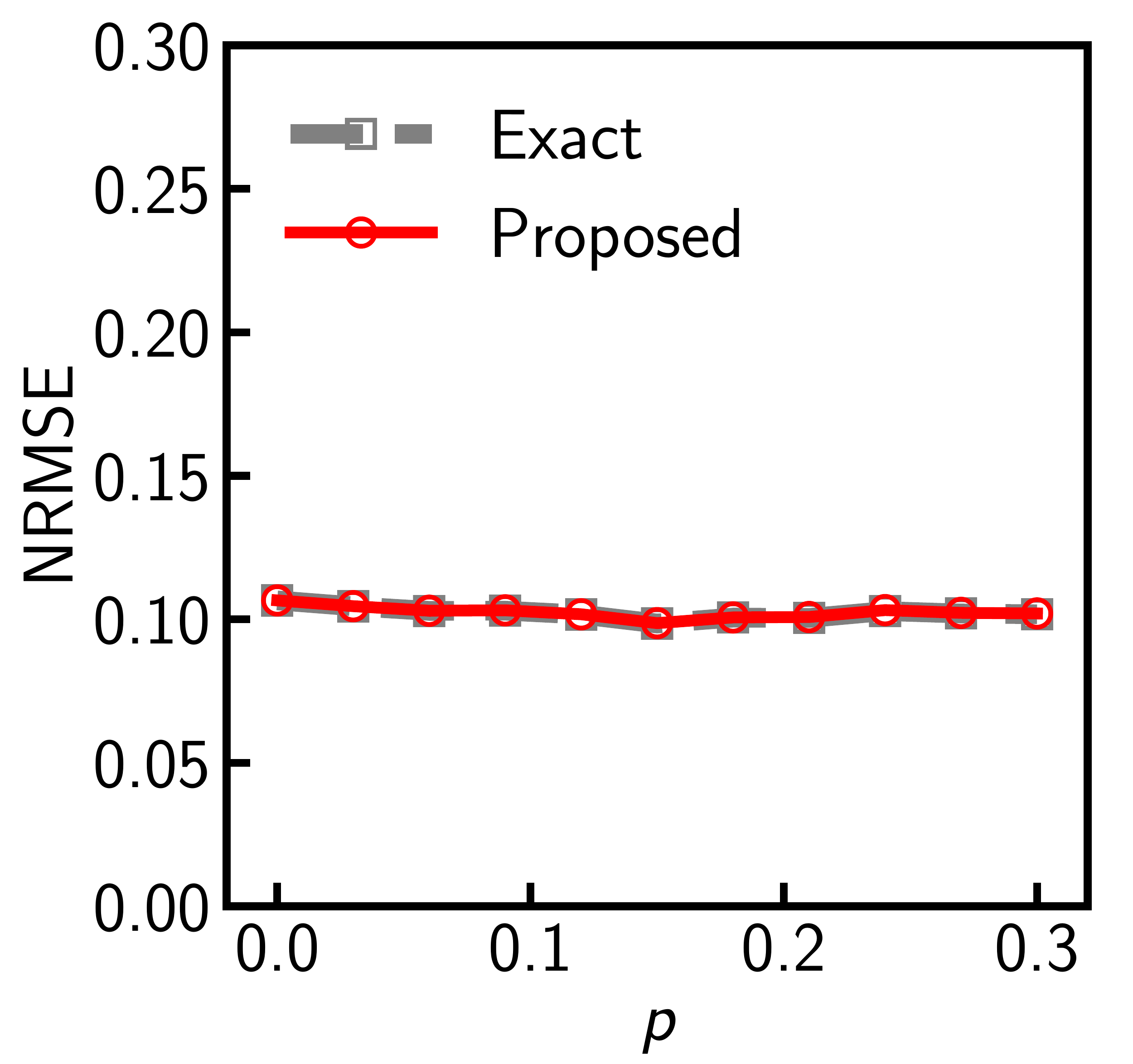}
          \\(a) NRMSEs of the proposed size estimator
      \label{pubd_accuracy}
        \end{center}
      \end{minipage}
      \begin{minipage}{0.48\hsize}
        \begin{center}
          \includegraphics[scale=\figscale]{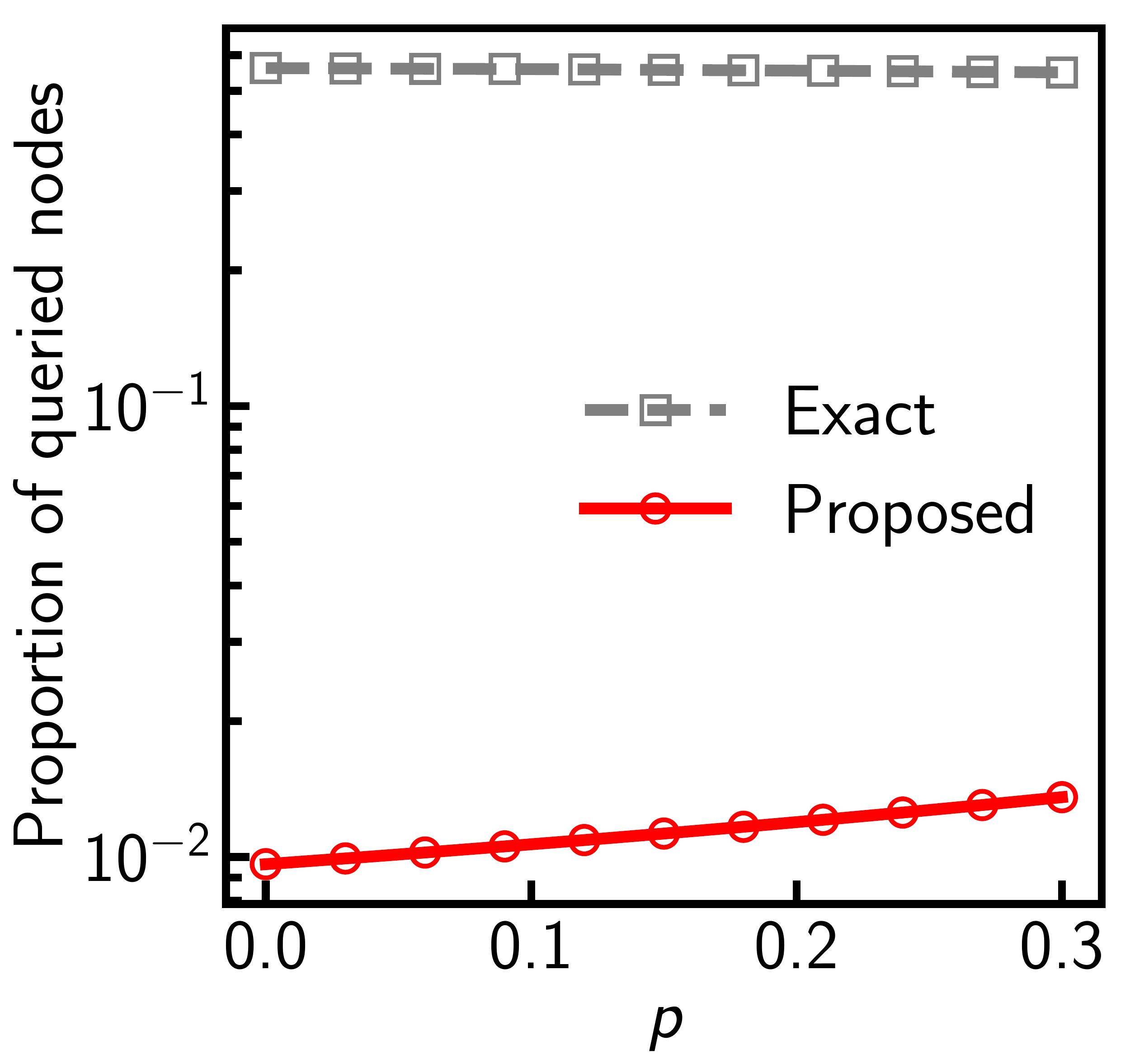}
          \\(b) Proportion of queried \\nodes 
      \label{pubd_query}
        \end{center}
      \end{minipage}
      \caption{Effects of the proposed calculation for public-degrees on Orkut (the hidden privacy model with 1\% sample size).}
      \label{pubd_compare}
\end{figure}

\subsection{Effectiveness of the Proposed Public-Degree Calculation} \label{effect_pubd}
We evaluate the proposed public-degree calculation for the hidden privacy model. 
The proposed size estimator requires the public-degree of each sample (see Section \ref{size_estimation}). 
Thus, we compare the size estimation accuracy and proportion of queried nodes of the exact and proposed methods. 
The exact method queries all neighbors of each sample to use the exact public-degree in weighting.

Figure \ref{pubd_compare} shows the experimental result with a 1\% sample size on Orkut.
The proposed approximation provides almost the same estimation accuracy as the exact method (see Figure \ref{pubd_compare} (a)).
The proposed method queries approximately 1\% nodes; however, the exact method queries over 50\% nodes which are much greater than 1\% sample size (see Figure \ref{pubd_compare} (b)).
The proposed method reduces the proportion of queried nodes by tens of percent with  almost the same accuracy on other datasets as well.

\subsection{Seed Selection from the Largest Public-Cluster} \label{seed}
In practical scenarios, we may require additional queries caused by restarting a random walk from another seed in the two cases: (1) when a selected seed is a private node and (2) when a selected public seed is on an isolated public-cluster.

We believe that the number of queries performed in each case is sufficiently small. 
For the case of (1), the proportion of private nodes is empirically smaller than that of public nodes, e.g., 27\% on Facebook \cite{catanese} and 34\% on Pokec \cite{takac}. 
Additionally, one query is enough to check the privacy label of a node.
For the case of (2), we believe that the number of queries is small under Assumption \ref{assumption_private}, because of the following two natures of graphs with degree distributions biased to low degrees \cite{albert}. 
First, most public nodes belong to the largest public-cluster. In our experiments, even if $p = 0.3$, 99.1\% , e.g., $\frac{0.694}{0.7}$ in Orkut, 91.4\%, e.g., $\frac{0.640}{0.7}$ in LiveJournal, 82.8\%, e.g., $\frac{0.580}{0.7}$ in Facebook and 76.7\%, e.g., $\frac{0.537}{0.7}$ in YouTube of public nodes belong to the largest public-cluster on average (see Figure \ref{cluster_size}). 
Therefore, a selected public seed will belong to the largest public-cluster with high probability. 
Second, the size of the isolated public-clusters is considerably smaller than that of the largest public-cluster. 
In our experiments, the average absolute size is only approximately one for every probability of $p$ (see Figure \ref{isolated_size}). 
Therefore, few queries are performed on isolated public-clusters. 

We believe that two real-world datasets support Assumption \ref{assumption_seed}.
There is only the largest public-cluster on Pokec; the case of (2) cannot occur.
The samples of real public Facebook users provide an estimated size of 657 million and contain one million unique samples; we believe that the samples were obtained from the largest public-cluster of Facebook as of October 2010. 

\section{Related Work}\label{related}
Crawling techniques that traverse neighbors are effective to sample graph data in OSNs where neighbor data is obtained through queries. 
The breadth-first search is particularly a basic sampling technique applied in early studies on analyzing the graph structure of OSNs \cite{ahn, gjoka_practical, gjoka_walking, mislove, wilson}. 
However, Gjoka et al., through the case study of crawling in Facebook, demonstrated that the breadth-first search with a small sample size leads to a significant sampling bias that is difficult to correct \cite{gjoka_practical, gjoka_walking}. 

Gjoka et al. designed a practical framework to obtain unbiased estimates of graph properties via a random walk-based sampling in social networks \cite{gjoka_practical, gjoka_walking}.  
Gjoka et al. clarified that the re-weighted random walk scheme, where each sample via random walk is weighted to correct the sampling bias, is effective to obtain unbiased estimates.
In this study, we have made the framework of Gjoka et al. more practical by addressing the effects of private nodes on the re-weighted random walk scheme for the first time.
Specifically, we have designed a random walk-based sampling algorithm considering private nodes and proposed the weighting methods to reduce both sampling bias and estimation errors due to private nodes.

The estimation algorithms based on the re-weighted random walk scheme have been studied for several properties \cite{chen, dasgupta, katzir_nodecc, katzir_node, nakajima, wang}. 
The targeted properties widely range from those of the graph such as the size and the average degree to those of the nodes like centralities.
We have designed the algorithms to accurately estimate the size and average degree of the whole graph including private nodes.
This study has focused on estimators for only the size and average degree; however, there is no particular reason to be limited to these. 
The goal of weighting methods to reduce both sampling bias and errors due to private nodes is shared in designing all of the random walk-based estimators for social networks. 
Then, the methodology presented in our study will help these designs.

Ye et al. investigated the effects of private nodes on general crawling methods and empirically argued that private nodes cause only a small reduction rate of the number of nodes and edges that can be sampled by crawling, under Assumptions \ref{assumption_public} and \ref{assumption_private} \cite{ye}. 
However, Ye et al. did not support the validity of assumptions.
In this study, we have focused on the effects of private nodes on random walk-based algorithms.
Moreover, we have validated the three assumptions through experiments using real-world datasets.

Kossinets experimentally discussed the effect of missing the nodes and edges on the properties of social networks \cite{kossinets}. 
Kossinets reported that as the proportion of randomly deleted nodes increases, errors of the size and average degree between the original graph and the remaining largest connected component increase. 
We have theoretically analyzed these errors and designed algorithms to improve them under certain assumptions and access models.
The important differences from the study of Kossinets are as follows: (i) we can find most private nodes in neighbors of public nodes on social networks under Assumption \ref{assumption_public} and (ii) we can reduce the errors due to private nodes by modifying the weighs for each sample obtained from the largest public-cluster based on Assumption \ref{assumption_private}.

\section{Conclusions}\label{conclusion}
We have designed re-weighted random walk algorithms considering private nodes for the first time.
The previous studies have not addressed the effects of private nodes on re-weighted random walk algorithms because private nodes prevent us from performing a simple random walk on a social graph. 
We have broken through this situation by making three assumptions. 
In particular, Assumption \ref{assumption_private} which is seemingly too simple has brought surprisingly reasonable results in experiments using real-world datasets.
Our study has opened the door to accurately estimate the whole graph topology of social networks including private nodes. 

There are mainly two future works. 
First, we plan to strengthen the validity of three assumptions. 
To further verify Assumptions \ref{assumption_public} and \ref{assumption_private}, we should additionally test the proposed estimators in real social networks including private nodes. 
To support Assumption \ref{assumption_seed} more, we will improve the sampling algorithm considering the case that the seed is on an isolated public-cluster. 
Second, we plan to design estimators considering private nodes for other properties, such as clustering coefficients \cite{katzir_nodecc} and motifs (or graphlets) \cite{chen, wang}.


\begin{acks}
We would like to thank Dr. Minas Gjoka for kind replies to our questions.
We would like to thank Prof. Naoto Miyoshi for helpful comments.
This work was supported by New Energy and Industrial Technology Development Organization (NEDO).
\end{acks}

\bibliographystyle{ACM-Reference-Format}
\bibliography{reference}

\newpage
\appendix

In the supplement, we show the pseudo code and proofs that could not be included in the main manuscript due to space limitations and the information for reproducing the experimental results.

\section{Pseudo-Code}

\begin{algorithm}[h]                     
\caption{Random walk in the hidden privacy model.}         
\label{RW_hidden}                          
\begin{algorithmic}[1]               
\REQUIRE seed $v_{x_1} \in C^*$, sample size $r$
\ENSURE sampling list $R$
\STATE $R \leftarrow$ empty list
\FOR{$k=1$ to $r$}
\STATE $d_{x_k} \leftarrow$ the degree of $v_{x_k}$
\STATE $\hat{d}_{x_k}^* \leftarrow 0$
\STATE append $(x_k,d_{x_k},\hat{d}_{x_k}^*)  \rightarrow R$
\IF{$v_{x_k}$ has been visited for the first time}
\STATE $a_{x_k} \leftarrow 0$
\STATE $b_{x_k} \leftarrow 0$
\ENDIF
\STATE flag $\leftarrow$ False
\WHILE{flag is False}
\STATE $v_{x_{k+1}} \leftarrow$ a neighbor of $v_{x_k}$ chosen randomly
\STATE $b_{x_k} \leftarrow b_{x_k} + 1$
\IF{$v_{x_{k+1}}$ is a public node}
\STATE $a_{x_k} \leftarrow a_{x_k} + 1$
\STATE flag $\leftarrow$ True
\ENDIF
\ENDWHILE
\ENDFOR
\FOR{$k=1$ to $r$}
\STATE $\hat{d}_{x_k}^* \leftarrow d_{x_k} \frac{a_{x_k}}{b_{x_k}}$
\ENDFOR
\RETURN{$R$}
\end{algorithmic}
\end{algorithm}

Algorithm \ref{RW_hidden} describes the designed random walk-based sampling using the proposed public-degree calculation in the hidden privacy model.
The process in lines from 11 to 18 will be successfully performed, because the $k$-th sample, $v_{x_k}$, has at least one public neighbor $v_{x_{k-1}}$ for $2 \leq k \leq r$ and $v_{x_1}$ is on the largest public-cluster. 
Additionally, it follows that $ b_{x_k}> 0$ for each sample $v_{x_k}$, because a neighbor selection is performed at least once for each sample.

\section{Proofs}
We show proofs of Lemmas \ref{distribution}, \ref{the_pubd_exp}, \ref{query_efficiency}, \ref{the_size_exp_naive}, \ref{error_size_naive}, \ref{the_size_exp_ours}, \ref{pubd_exp}, \ref{the_aved_exp_naive}, \ref{error_aved_naive}, \ref{the_aved_exp_ours}, Proposition \ref{p0_size}, Theorems \ref{error_size_ours}, \ref{error_aved_ours}, and Corollary \ref{relative_size}.

\subsection{Proof of Lemma \ref{distribution}}
\begin{proof}
First, it holds $Pr[x_r = i] = 0$ for each node $v_i \in V \backslash V^*$, because we cannot reach nodes not in $V^*$. Then, for each node $v_i \in V^*$, we show that $Pr[x_r = i]$ converges to $\frac{d_i^*}{D^*}$ after many steps. The designed random walk started from the seed on $C^*$ has the transition probability matrix $\bm{P} = \{P_{i,j}\}_{v_i,v_j \in V^*}$, where $P_{i,j}$ is $\frac{1}{d_i^*}$, if $(v_i, v_j) \in E^*$, otherwise, 0. $\bm{P}$ is ergodic because $C^*$ is a connected graph, and hence, the stationary distribution uniquely exists because of the Theorem \ref{ergodic}. $\{p_i\}_{v_i \in V^*}$ satisfies the definition of the stationary distribution. The probability $Pr[x_r = i]$ converges to the corresponding stationary distribution after many steps.
\end{proof}

\subsection{Proof of Lemma \ref{the_pubd_exp}}
\begin{proof}
Let $\hat{X}_k(l)$ denote a random variable which returns 1 if a public neighbor of $v_{x_k}$ is selected at the $l$-th trial, otherwise 0, where $1 \leq l \leq b_{x_k}$ and it holds $\sum_{l=1}^{b_{x_k}} \hat{X}_k(l) = a_{x_k}$. It holds $Pr[\hat{X}_k(l) = 1] = \frac{d_{x_k}^*}{d_{x_k}}$ because $\hat{X}_k(l)$ follows a Bernoulli distribution for each $l$. Then, it holds $E[\hat{d}_{x_k}^*] = d_{x_k} \frac{d_{x_k}^*}{d_{x_k}} = d_{x_k}^*$. Because $\{\hat{X}_k(l)\}_{l = 1}^{b_{x_k}}$ is a process of independent and identically distributed trials, $\hat{d}_{x_k}^*$ converges to its expectation after many steps of the designed random walk because of the law of large numbers.
\end{proof}

\subsection{Proof of Lemma \ref{query_efficiency}}
\begin{proof}
It holds $Q(k) = d_{x_k}$, because the exact method queries all neighbors of each sample. Therefore, we have 
\begin{align*}
E[Q] = E[Q(k)] = \sum_{v_i \in V^*} {\frac{d_i^*}{D^*}} E[Q(k) | x_k = i] =  \frac{\sum_{v_i \in V^*} d_i^* d_i}{D^*}.
\end{align*}
The first and second equations hold because of the linearity of expectation and the law of total expectation.

$\hat{Q}(k)$ follows the geometric distribution with success probability $\frac{d_{x_k}^*}{d_{x_k}}$, because the proposed method randomly queries  neighbors of $v_{x_k}$ until the public neighbor is firstly selected. Thus, it holds $E[\hat{Q}(k)] = \frac{d_{x_k}}{d_{x_k}^*}$. Then, we have
\begin{align*}
E[\hat{Q}] = E[\hat{Q}(k)] = \sum_{v_i \in V^*} {\frac{d_i^*}{D^*}} \frac{d_i}{d_i^*} = \frac{\sum_{v_i \in V^*} d_i}{D^*}.
\end{align*}
Lemma \ref{query_efficiency} holds because of the above equations.
\end{proof}

\subsection{Proof of Lemma \ref{the_size_exp_naive}}
\begin{proof}
First, we calculate the expectation of $\Phi^{NC}_{n}$:
\begin{align*}
E\left[\Phi^{NC}_{n}\right] &= E[\phi_{k,l}] = \sum_{v_i \in V^*} {\left(\frac{d_i^*}{D^*}\right)^2}.
\end{align*}
The first equation holds because of the linearity of expectation. The second equation holds because $v_{x_k}$ and $v_{x_l}$ are independently sampled with the probability $\frac{d_i^*}{D^*}$. Then, we obtain 
\begin{align*}
E\left[\Psi^{NC}_{n}\right] &= E\left[\frac{d_{x_k}^*}{d_{x_l}^*}\right] = \sum_{v_i \in V^*} {\sum_{v_j \in V^*} {\frac{d_i^*}{d_j^*}\frac{d_j^*}{D^*}\frac{d_i^*}{D^*}}} = n^* \sum_{v_i \in V^*} {\left(\frac{d_i^*}{D^*}\right)^2}.
\end{align*}
We conclude that $n^{NC}$ converges to $n^*$, because $\Phi^{NC}_{n}$ and $\Psi^{NC}_{n}$ intuitively converge to their respective expectations.
\end{proof}

\subsection{Proof of Lemma \ref{error_size_naive}}
\begin{proof}
We define the random variable $X_n(i) = \bm{1}_{V^*}(v_i)$ for each node $v_i \in V$. It holds $n^* = \sum_{v_i \in V} X_n(i)$. The expected value of $n^*$ regarding $\mathcal{L}_{pri}$ is given by:
\begin{align*}
  E_{pri}[n^*] &= \sum_{v_i \in V}E_{pri}[X_n (i)] = \sum_{v_i \in V} Pr[v_i \in V^*] = (1-p)n.
\end{align*}
The first and second equations hold because of the linearity of the expected value and the law of total expectation.
\end{proof}

\subsection{Proof of Lemma \ref{the_size_exp_ours}}
\begin{proof}
As with the proof of the Lemma \ref{the_size_exp_naive}, we have
\begin{align*}
E\left[\hat{\Psi}_{n}\right] = E\left[\frac{d_{x_k}}{d_{x_l}^*}\right] = n^* \sum_{v_i \in V^*} {\frac{d_i^* d_i}{(D^*)^2}}.
\end{align*}
Because $\Phi^{NC}_{n}$ and $\hat{\Psi}_{n}$ intuitively converge to the expected values respectively, $\hat{n}$ converges to $\tilde{n}$.
\end{proof}

\subsection{Proof of Proposition \ref{p0_size}}
\begin{proof}
When there are no private nodes on $G$, we have $V^* = V$ and $d_i^* = d_i$ for each node $v_i \in V^*$, because $C^*$ is equivalent to $G$. Thus, it holds $n^{NC} = \hat{n}$ from the definitions. It holds $n^* = \tilde{n} = n$ because of Lemmas \ref{the_size_exp_naive} and \ref{the_size_exp_ours}.
\end{proof}

\subsection{Proof of Lemma \ref{pubd_exp}}
\begin{proof}
$d_i^*$ follows the binomial distribution with parameters $d_i$ and $1-p$ regarding $\mathcal{L}_{pri}$, because each neighbor of $v_i$ independently becomes a public node with probability $1-p$.
\end{proof}

\subsection{Proof of Theorem \ref{error_size_ours}}
\begin{proof}
We define the random variables $\hat{X}_n (i) = d_i^* d_i \bm{1}_{V^*}(v_i)$ and $\hat{Y}_n (i) = (d_i^*)^2 \bm{1}_{V^*}(v_i)$ for each node $v_i \in V$. Also, let $\hat{X} = \sum_{v_i \in V^*} d_i^* d_i$ and $\hat{Y} = \sum_{v_i \in V^*} (d_i^*)^2$. It holds $\hat{X} = \sum_{v_i \in V} \hat{X}_n (i)$ and $\hat{Y} = \sum_{v_i \in V} \hat{Y}_n (i)$. We obtain the expectation of $\hat{X}$ regarding $\mathcal{L}_{pri}$:
\begin{align*}
  E_{pri}[\hat{X}] = \sum_{v_i \in V}Pr[v_i \in V^*]E_{pri}[d_i^* d_i] = (1-p)^2 \sum_{v_i \in V}(d_i)^2.
\end{align*}
The second equation hold because of Lemma \ref{pubd_exp}. We note that the degree $d_i$ is a constant regarding $\mathcal{L}_{pri}$. Similarly, the expectation of $\hat{Y}$ regarding $\mathcal{L}_{pri}$ can be obtained as follows:
\begin{align*}
  E_{pri} [\hat{Y}] &= \sum_{v_i \in V}Pr[v_i \in V^*]E_{pri}[(d_i^*)^2] \\
  &= (1-p)^2 \sum_{v_i \in V} d_i [(1-p)d_i + p].
\end{align*}
Theorem \ref{error_size_ours} holds because of above equations and Lemma \ref{error_size_naive}.
\end{proof}

\subsection{Proof of Corollary \ref{relative_size}}
\begin{proof}
From Lemma \ref{error_size_naive} and Theorem \ref{error_size_ours}, it is sufficient to show that $|1-\alpha_p| \leq p$. First, we show that $1- \alpha_p \geq 0$:
\begin{align*}
  1 - \alpha_p = \frac{p\sum_{v_i \in V} d_i}{\sum_{v_i \in V} d_i [(1-p)d_i + p]} \geq 0.
\end{align*}
The inequality holds because it holds $d_i \geq 1$ for each node $v_i \in V^*$ and $0 \leq p \leq 1$. It holds equality when $p = 0$. Therefore, it is enough to show that $1- \alpha_p \leq p$. We have 
\begin{align*}
  1 - \alpha_p - p = -(1-p)p \frac{\sum_{v_i \in V} [d_i (d_i - 1)]}{\sum_{v_i \in V} d_i [(1-p)d_i + p]} \leq 0.
\end{align*}
The inequality holds since it holds $d_i \geq 1$ for each node $v_i \in V^*$ and $0 \leq p \leq 1$.
\end{proof}

\subsection{Proof of Lemma \ref{the_aved_exp_naive}}
\begin{proof}
Let $\Phi_{avg}^{Smooth}$ be the weighted average of $\frac{1}{d_{x_k}^*}$ for samples. We calculate the expectation of $\Phi_{avg}^{Smooth}$ as follows:
\begin{align*}
E\left[\Phi_{avg}^{Smooth}\right] &= E\left[\frac{1}{d_{x_k}^*}\right] = \sum_{v_i \in V^*} \frac{d_i^*}{D^*} \frac{1}{d_i^*} = \frac{1}{d_{avg}^*}.
\end{align*}
Because $\Phi_{avg}^{Smooth}$ converges to the expected value after many steps because of Theorem \ref{SLLN}, $d_{avg}^{Smooth}$ converges to $d_{avg}^*$.
\end{proof}

\subsection{Proof of Lemma \ref{error_aved_naive}}
\begin{proof}
We define the random variable $X_{avg} (i) = d_i^* \bm{1}_{V^*}(v_i)$ for each node $v_i \in V$. It holds $D^* = \sum_{v_i \in V} X_{avg}(i)$. The expected value regarding $\mathcal{L}_{pri}$ of $D^*$ can be derived:
\begin{align*}
  E_{pri}[D^*] &= \sum_{v_i \in V}Pr[v_i \in V^*]E_{pri}[d_i^*] = (1-p)^2 D.
\end{align*}
Consequently, it follows that Lemma \ref{error_aved_naive} holds because of the above equation and Lemma \ref{error_size_naive}.
\end{proof}

\subsection{Proof of Lemma \ref{the_aved_exp_ours}}
\begin{proof}
Let $\hat{\Phi}_{avg} = \frac{1}{r} \sum_{k=1}^r {\frac{1}{d_{x_k}}}$ be the weighted average of $\frac{1}{d_{x_k}}$ for samples. Then, we have:
\begin{align*}
E[\hat{\Phi}_{avg}] = E\left[\frac{1}{d_{x_k}}\right] = \sum_{v_i \in V^*} \frac{d_i^*}{D^*} \frac{1}{d_i}= \frac{1}{\tilde{d}_{avg}}.
\end{align*}
Since $\hat{\Phi}_{avg}$ converges to the expected value after many steps because of Theorem \ref{SLLN}, $\hat{d}_{avg}$ converges to $\tilde{d}_{avg}$.
\end{proof}

\subsection{Proof of Theorem \ref{error_aved_ours}}
\begin{proof}
We define the random variable $\hat{X}_{avg} (i) = \frac{d_i^*}{d_i} \bm{1}_{V^*}(v_i)$ for each node $v_i \in V$. Also, let $\hat{X}_{avg} = \sum_{v_i \in V^*} \frac{d_i^*}{d_i}$. It holds $\hat{X}_{avg} = \sum_{v_i \in V}\hat{X}_{avg} (i)$. 
We obtain the expectation of $\hat{X}_{avg}$ regarding $\mathcal{L}_{pri}$:
\begin{align*}
  E_{pri}[\hat{X}_{avg}] &= \sum_{v_i \in V}Pr[v_i \in V^*]E_{pri}\left[\frac{d_i^*}{d_i}\right] = (1-p)^2 n.
\end{align*}
Using the fact that $E_{pri}[D^*]=(1-p)^2 D$ and the above equation, we have that the Theorem \ref{error_aved_ours} holds.
\end{proof}

\section{Additional Information for Reproducing Experimental Results}
All the original datasets are publicly available as follows: \\
{\bf YouTube:} \url{http://konect.uni-koblenz.de/networks/com-youtube}\\
{\bf Pokec:} \url{http://snap.stanford.edu/data/soc-Pokec.html}\\
{\bf Orkut:} \url{https://snap.stanford.edu/data/com-Orkut.html}\\
{\bf Facebook:} \url{http://networkrepository.com/socfb-A-anon.php}\\
{\bf LiveJournal:} \url{https://snap.stanford.edu/data/com-LiveJournal.html}\\
{\bf Real public Facebook user samples:} \url{http://odysseas.calit2.uci.edu/doku.php/public:osn_datasets#facebook_weighted_random_walks}\\

The datasets and source code that were used to generate the results in this paper are available at \url{https://github.com/kazuibasou/KDD2020}.

\end{document}